\documentclass[
    reprint,
    superscriptaddress,
    a4paper,
    amsmath,amssymb,
    prb,
]{revtex4-2}
\usepackage{graphicx, color} %
\graphicspath{{fig/}}

\usepackage[dvipsnames]{xcolor}
\definecolor{APSBlue}{HTML}{2e3092}
\usepackage[
    colorlinks,
    citecolor=APSBlue,
    linkcolor=APSBlue,
    urlcolor=APSBlue,
]{hyperref}
\usepackage{lipsum}
\usepackage{comment}

\usepackage{textcomp}
\usepackage{graphicx}
\usepackage{fixmath}
\usepackage{xcolor}

\usepackage{stackengine}
\setstackgap{S}{1pt}

\usepackage{physics}

\newcommand{\iu}{\mathrm{i}\mkern1mu}
\newcommand{\eu}{\mathrm{e}\mkern1mu}

\usepackage{xspace}  
\usepackage{chemformula}

\newcommand{\affilANU}{Nonlinear Physics Center, Research School of Physics, Australian National University, Canberra ACT 2601, Australia}

\newcommand{\affilHarbin}{Ministry of Industry and Information Technology Key Lab of Micro-Nano Optoelectronic Information System, Guangdong Provincial Key Laboratory of Semiconductor
Optoelectronic Materials and Intelligent Photonic Systems, Harbin Institute of Technology, Shenzhen 518055, P.R. China}

\newcommand{\affilShubnikov}{ Shubnikov Institute of Crystallography, NRC “Kurchatov Institute”, Moscow 119333, Russia}

\newcommand{\affilMEPhI}{National Research Nuclear University MEPhI (Moscow Engineering Physics Institute), Moscow 115409, Russia}

\begin{document}

\title{Chiral Dichroism in Resonant Metasurfaces with Monoclinic Lattices}

\author{Ivan Toftul}
\email{ivan.toftul@anu.edu.au}
\affiliation{\affilANU}

\author{Pavel Tonkaev}
\affiliation{\affilANU}

\author{Kirill Koshelev}
\affiliation{\affilANU}

\author{Fangxing Lai}
\affiliation{\affilHarbin}

\author{Qinghai Song}
\affiliation{\affilHarbin}

\author{Maxim Gorkunov}
\affiliation{\affilShubnikov}
\affiliation{\affilMEPhI}

\author{Yuri Kivshar}
\email{yuri.kivshar@anu.edu.au}
\affiliation{\affilANU}

\begin{abstract}
 We demonstrate that chiral response can be achieved in resonant metasurfaces with a monoclinic lattice symmetry (the so-called Bravais {\it oblique lattices}) where the mirror symmetry is broken by the lattice asymmetry and also by a substrate, whereas each individual meta-atom remains fully {\it achiral}. We describe the underlying physics by introducing {\it a mode chirality parameter} as a quantitative measure of the lattice chiral eigenmodes. We confirm experimentally selective linear and nonlinear chiral interaction of resonant silicon metasurfaces with circularly polarized light.
\end{abstract}

\maketitle

\textit{Introduction.}---Chiral metasurfaces promise to become invaluable tools capable of emission, selective detection and transformation of chiral light~\cite{Khaliq2023AOM} required for chiral biosensing~\cite{Avalos-Ovando2022Jul}, chiral photochemistry~\cite{He2018NC} and chiral quantum optics~\cite{Lodahl2017N}. Simplifying chiral metastructures to the utmost possible extent is critical for facilitating their future mass production for all versatile practical implementations. Traditionally, chiral response of a metasurface is achieved by engineering the \textit{complex shape} of meta-atoms. It is done by shaping metallic meta-atoms~\cite{Papakostas2003Mar,Ren2012May} or by creating multilayer~\cite{Tanaka2020Nov} or single-layer asymmetric dielectric structures~\cite{Koshelev2023ACSPhot}. 
Another option is to utilize the \textit{rotation} of meta-atoms with respect to the overall metasurface lattice.
The later possibility was discussed theoretically~\cite{Movsesyan2022Jul,Avalos-Ovando2022Jul} and demonstrated  experimentally~\cite{Gryb2023Oct} very recently.
However, these methods generally require meta-atoms with inherent noncontinuous rotational symmetry (i.e., lacking infinite rotational symmetry, $C_{\infty}$), which generally complicates the design and fabrication.
Moir\'e chiral photonic metasurfaces can also give a chiral response~\cite{Lyu2023LPR,Wu2018Nanoscale} but that would require a challenging fabrication procedure as well.

\begin{figure*}
    \centering
    \includegraphics[width=1\linewidth]{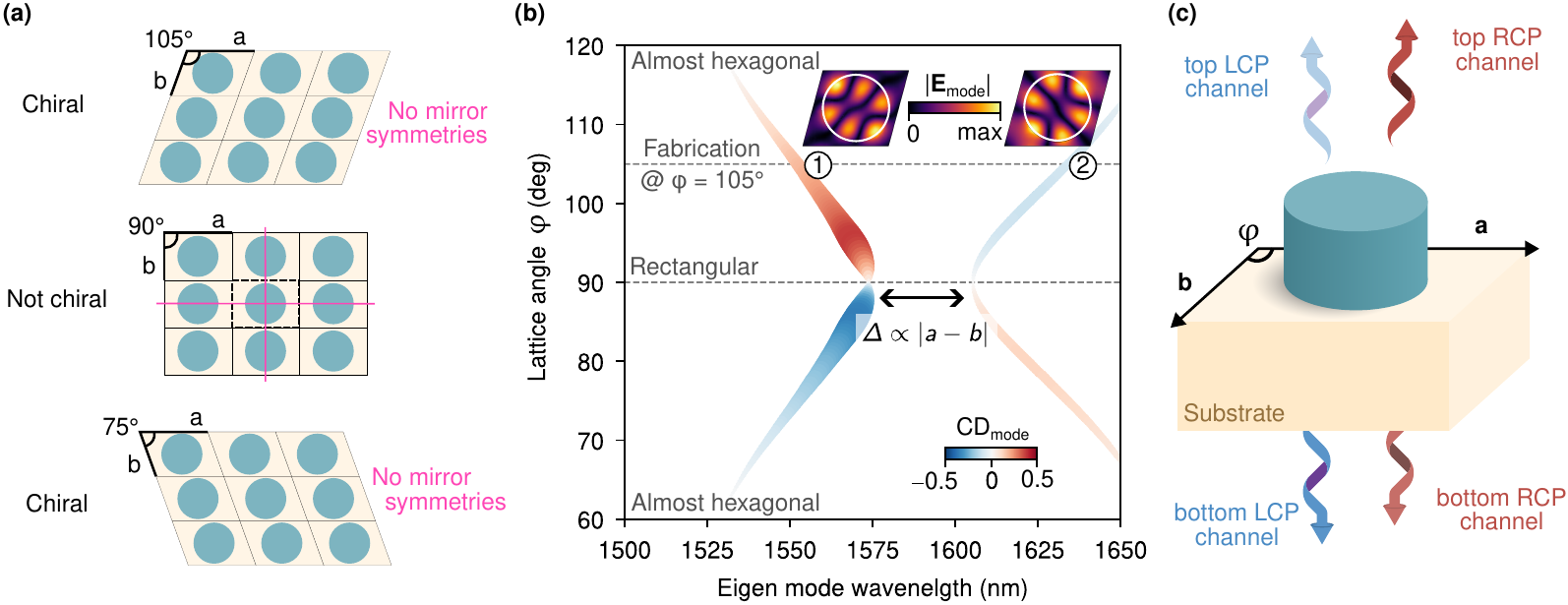}
    \caption{
    \textbf{The concept of chiral metasurfaces build of achiral meta-atoms}. 
    \textbf{(a)} Absence of mirror symmetry planes makes the metasurface geometrically chiral. One of the possible ways is to use monoclinic lattice arrangement of simple-shaped achiral meta-atoms, such as cylinders.
    \textbf{(b)} Metasurface eigenmodes acquire chirality once the lattice angle is not equal to $90^{\circ}$, i.e., the lattice is not rectangular.
    To avoid mode degeneracy at $\varphi = 90^{\circ}$, the lattice constants are taken unequal, $a\neq b$, and we observe a typical anti-crossing behaviour. 
    At the angles of $\varphi = 60^{\circ}$ and $120^{\circ}$ the lattice becomes very close to hexagonal and the mode chirality vanishes as this lattice has the highest number of in-plane mirror symmetry planes.
    \textbf{(c)} Schematic of a chiral mode differently coupled with RCP and LCP channels. 
    }
    \label{fig:concept}
\end{figure*}

While extrinsic chirality manifested for obliquely incident beams is a well-established phenomenon~\cite{Plum2009May,Asefa2023Sep}, it often necessitates complex setups and may not be ideal for practical applications. 
Here we propose how to %
achieve both linear and nonlinear intrinsic chiral optical responses at \textit{normal} incidence utilizing ultimately symmetric achiral cylindrical meta-atoms arranged in monoclinic (in terms of 2D Bravais lattice types) lattices. 
The approach opens a broad avenue for manipulating chiral light-matter interactions, and it will substantially facilitate many useful practical applications due to the ultimate simplicity of required metastructures.

\begingroup
\squeezetable
\begin{table}[b]
\caption{\label{tab:summary}Summary of metasurface chiral 
transmission characteristics, approximations at the resonance in the pump frequency range, and symmetry requirements.}
\begin{ruledtabular}
\begin{tabular}{lccc}
Characteristics                                                     & Definitions & \begin{tabular}[c]{@{}c@{}} Res. approx.\\ $\omega \approx \omega^{\prime}_n$\end{tabular}   & $\neq 0$ if  \\ \hline
\begin{tabular}[c]{@{}l@{}}Linear CD \\ (side independent)\end{tabular}                                                        &   Eq.~\eqref{eq:CDco}         &     Eq.~\eqref{eq:CDmode}                          & \begin{tabular}[c]{@{}c@{}}true chiral \\ (no mirror sym.)\end{tabular} \vspace{0.4cm} \\
\begin{tabular}[c]{@{}l@{}}Nonlinear CD \\ (pump from air)\end{tabular} &      Eq.~\eqref{eq:NCDtot}       &   Eq~\eqref{eq:NCDn}   &   no in-plane sym.                        \vspace{0.4cm} \\
\begin{tabular}[c]{@{}l@{}}Nonlinear CD \\ (pump from subs.)\end{tabular} &       \begin{tabular}[c]{@{}c@{}}Eq~\eqref{eq:NCDtot} \\ ($I^{(3\omega)} \to I^{\prime (3\omega)}$)\end{tabular}     &    \begin{tabular}[c]{@{}c@{}}Eq~\eqref{eq:NCDn} \\ ($m \to m^{\prime}$)\end{tabular}   &      no in-plane sym.
\end{tabular}
\end{ruledtabular}
\end{table}
\twocolumngrid
\endgroup

\textit{Concept and theoretical prediction.}---Chirality is a geometric property according to which an object is mirror-asymmetric, or, equivalently, different from its image seen in a mirror, irrespective of orientation~\cite{Caloz2020Feb}.
The metasurface made of rotationally symmetric meta-atoms distributed in a monoclinic lattice on a substrate is chiral, Fig.~\ref{fig:concept}{(a)}.
This mirror asymmetry can be optically detected by exposing the object of interest to right circularly polarized (RCP) light and left circularly polarized (LCP) light.
The conventional characteristic is \textit{circular dichroism} (CD) which is originally defined as a difference in \textit{absorption} of RCP and LCP light, and this definition is widely adopted in chemistry and biology~\cite{fasman2013circular,Kobayashi2011Nov,Baase1979Feb}. 
In the context of dielectric metasurfaces, the CD definition becomes less related to absorption, as drastically different response to LCP and RCP light is possible in lossless structures~\cite{Hu2017Jan,Gorkunov2020Aug,Kim2020Aug}.
Moreover, in metasurfaces of different point symmetry, the asymmetry of right-to-right and left-to-left \textit{copolarized transmission} can be accompanied by the right-to-left  and left-to-right \textit{cross-polarized transmission} 
asymmetry \cite{Koshelev2024JOPT}. 
As, nevertheless, a reciprocal achiral structure must have symmetric co-polarized  transmission, the copolarized CD defined as
\begin{equation}
    \mathrm{CD}_{\text{co}} = \frac{\abs{t_{\text{RR}}}^2 - \abs{t_{\text{LL}}}^2}{\abs{t_{\text{RR}}}^2 + \abs{t_{\text{LL}}}^2} = \frac{\abs{t^{\prime}_{\text{RR}}}^2 - \abs{t^{\prime}_{\text{LL}}}^2}{\abs{t^{\prime}_{\text{RR}}}^2 + \abs{t^{\prime}_{\text{LL}}}^2}
    \label{eq:CDco}
\end{equation}
remains an explicit indicator of the geometric chirality. 
It varies between $-1$ and $+1$, and here $t_{\text{RR}}$ and $t_{\text{LL}}$ are the complex transmission coefficients in the circular basis, with the first and last indexes denoting the output and input polarizations, correspondingly; primed transmission coefficients describe the case of excitation from the opposite side, i.e., from a substrate.
$\mathrm{CD}_{\text{co}}$ in Eq.~\eqref{eq:CDco} is independent of the input direction as long the structure consists of electromagnetically reciprocal materials. 
The Lorentz reciprocity imposes different limitations on the reflection coefficients, as it requires the equality $r_{\text{RL}}=r_{\text{LR}}$ at the normal incidence, regardless of the metasurface symmetry~\cite{Koshelev2024JOPT,Gorkunov2024}. 
The summary of metasurface chiral transmission characteristics studied in this Letter is in Table.~\ref{tab:summary}.

While a mirror asymmetric metasurface is allowed to have $\mathrm{CD}_{\text{co}} \neq 0$, symmetry breaking does not guarantee that CD reaches high values. This requires careful engineering of metasurface structure, which, in particular, determines the spatial structure of metasurface photonic eigenmodes. 
Particularly attractive are high quality factor (high-$Q$) modes selectively coupled to an RCP or an LCP radiation channel. Chiral bound states in the continuum are representative examples of such modes~\cite{Gorkunov2020Aug,Overvig2021Feb,kuhner2023}.

In this Letter, we adopt and develop a highly efficient predictive method that allows designing strongly chiral metasurfaces based only on the fast eigenmodes calculations rather than on computationally expensive full numeric transmission calculations.
We assume that in the vicinity of the resonance, the metasurface response is determined by a single resonant mode. By applying the coupled mode theory, we can quantify the eigenmode contribution to the chiral optical properties, and hence the potential values of $\mathrm{CD}_{\text{co}}$. It is particularly useful to introduce the \textit{mode circular dichroism}, $\mathrm{CD}_{\text{mode},n}$, as~\cite{SM}
\begin{equation}
    \mathrm{CD}_{\text{mode},n} = \frac{\abs{m^{\prime}_{n \text{R}}m_{n\text{R}}}^2 - \abs{m^{\prime}_{n\text{L}}m_{n\text{L}}}^2}{\abs{m^{\prime}_{n\text{R}}m_{n\text{R}}}^2 + \abs{m^{\prime}_{n\text{L}}m_{n\text{L}}}^2}
    \label{eq:CDmode}
\end{equation}
where the parameters of the mode coupling to RCP and LCP waves are introduced as~\cite{Weiss2018PRB,Gorkunov2020Aug,koshelev2022PhDthesis,Koshelev2020Science}:
\begin{equation}
    m_{n\text{R,L}} = A_n \int \limits_{\text{meta}} \Delta \varepsilon (\vb{r}) \vb{E}_n (\omega_n, \vb{r}) \cdot \vb{E}_{\text{bg}}^{(\mathrm{R}, \mathrm{L})} (\omega_n, \vb{r}) \dd^{3} \vb{r}
    \label{eq:m}
\end{equation}
with the eigenmode electric field $\vb{E}_n$ and the background electric field of RCP or LCP wave $\vb{E}_{\text{bg}}^{(\mathrm{R},\mathrm{L})}$ including the Fresnel fields reflected and transmitted by the substrate without the metasurface~\cite[Sec.~VI]{Weiss2018PRB,note_one}. 
$\Delta \varepsilon (\vb{r})$ denotes the difference between the permittivities of the metasurface and the background which in our case includes the substrate.
The primed coupling parameters in \eqref{eq:CDmode} describe the coupling to the waves incident on the back metasurface side. See also Fig.~\ref{fig:concept}{(c)} for the illustration of the outgoing channels.
The $A_n$ is the normalization coefficient.  
A symmetric magnetic counterpart of Eq.~\eqref{eq:m} is absent due to the nonmagnetic nature of the materials, i.e., $\Delta \mu (\vb{r}) =0$.
Coefficient $A_n$ is unique for each mode, which includes the eigenmode normalization, perturbation of the permittivity, and other constants (see Appendix of Ref.~\cite{koshelev2022PhDthesis}) but we do not provide its explicit form as it cancels out in Eq.~\eqref{eq:CDmode}.

\begin{figure*}
    \centering
    \includegraphics[width=0.85\linewidth]{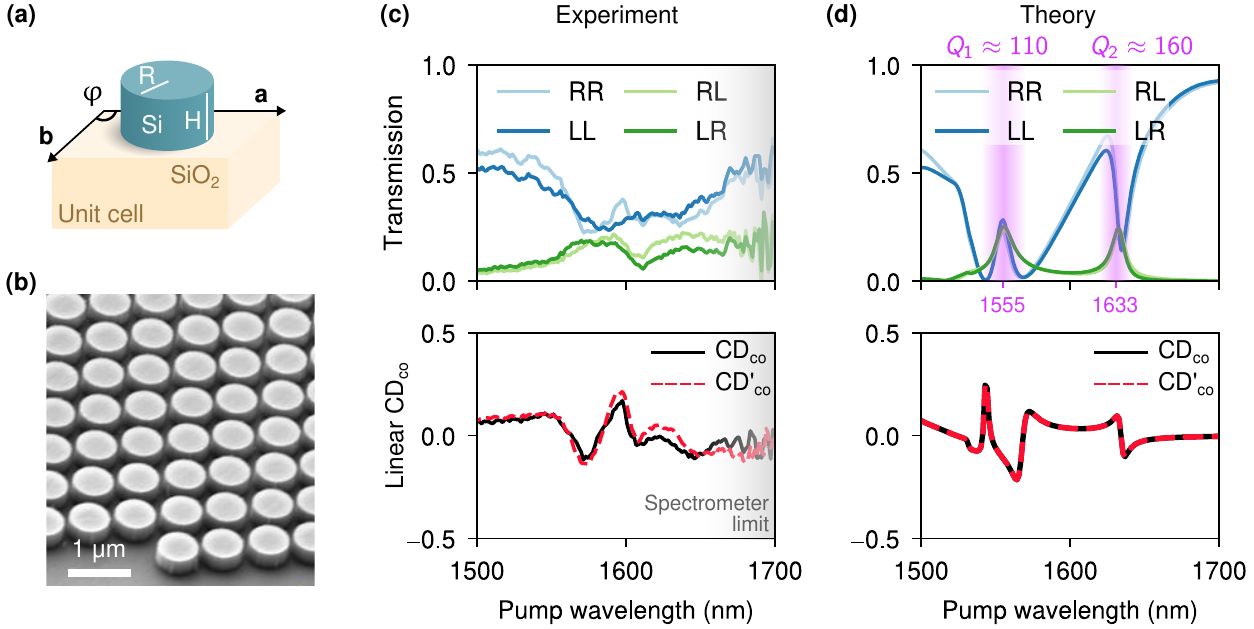}
    \caption{
    \textbf{Linear optical properties}. 
    \textbf{(a)} Schematic representation of the metasurface unit cell with all geometrical parameters.
    \textbf{(b)} Scanning electron microscope image of a fragment of fabricated sample.
    \textbf{(c)} and \textbf{(d)} are the experimental and theoretical spectra of linear transmission coefficients and circular dichroism.
    Theory predict the position of two modes at $\lambda_{1}^{\text{res}} = 1555$~nm and at $\lambda_{2}^{\text{res}} = 1633$~nm with quality factors of $Q_1 \approx 110$ and $Q_2 \approx 160$, respectively. $\mathrm{CD}_{\text{co}}$ and $\mathrm{CD}^{\prime}_{\text{co}}$ denote co-polarized CD of transmission of light incident from the air and from the substrate correspondingly. See simulation details in~\cite{SM}.
    }
    \label{fig:linear}
\end{figure*}

We start our design with the Si cylinders in a rectangular lattice [$a\neq b$, $\varphi = 90^{\circ}$, Fig.~\ref{fig:concept}{(a)}] placed on a substrate made of SiO$_2$. 
The two eigenmodes of interest, which electric field distribution is shown in the inset of Fig.~\ref{fig:concept}{(b)}, are coupled to each other. The degeneracy is removed by making the lattice rectangular rather than square, i.e., spectral distance between the eigenmodes increases as $\Delta \propto |a - b|$, where $a$ and $b$ stand for the lattice spacings~\cite{SM}. The out-of-plane mirror symmetry is {\it intrinsically broken} by the substrate. Next, we break all in-plane mirror symmetries by making the lattice monoclinic. The optimal monoclinic lattice angle, $\varphi$, can be obtained by examining Fig.~\ref{fig:concept}{(b)}, where we plot the mode circular dichroism~\eqref{eq:CDmode} for these modes.
As expected, for $\varphi = 90^{\circ}$, the mode circular dichroism is zero, $\mathrm{CD}_{\text{mode}} = 0$, as the lattice is rectangular and has 2 in-plane mirror symmetries. For lattice angles $\varphi = 60^{\circ}$ or $\varphi = 120^{\circ}$ we obtain $\mathrm{CD}_{\text{mode}} \approx 0$ as the metasurface lattice approaches a hexagonal lattice, which has 6 in-plane mirror symmetries.
For intermediate $\varphi$ values, there are no mirror symmetries which results in nonzero mode chirality parameter values, $\mathrm{CD}_{\text{mode}} \neq 0$. 
Metasurfaces with $\varphi = 90^{\circ} - \Delta \varphi$   and $\varphi = 90^{\circ} + \Delta \varphi$ are enantiomers, where $\Delta \varphi$ ranges from $0^{\circ}$ to $30^{\circ}$.

Mode circular dichroism \eqref{eq:CDmode} is not directly related to the circular dichroism~\eqref{eq:CDco}, and  $\mathrm{CD}_{\text{mode}} \approx \mathrm{CD}_{\text{co}}\big|_{\omega = \Re(\omega_n)}$ only in the vicinity of the resonance when contribution of all other modes is negligibly small and the background achiral transmission is absent~\cite{SM}.

\textit{Experimental realization.}---To support our theoretical claim we fabricate and characterize such design, Figs.~\ref{fig:linear}(b),~\ref{fig:linear}(d).  The design parameters were $a = 1100$~nm, $b=1000$~nm, $R = 430$~nm, $H = 400$~nm, and $\varphi = 75^{\circ}$.
The qualitative compliance is observed, however the measured spectra are not perfectly matched with the theory due to the finite size of the real structure, imperfections during fabrication, and Gaussian beam excitation in contrast to a plane wave in the theoretical calculations. See details in~\cite{SM}.

\begin{figure*}
    \centering
    \includegraphics[width=0.98\linewidth]{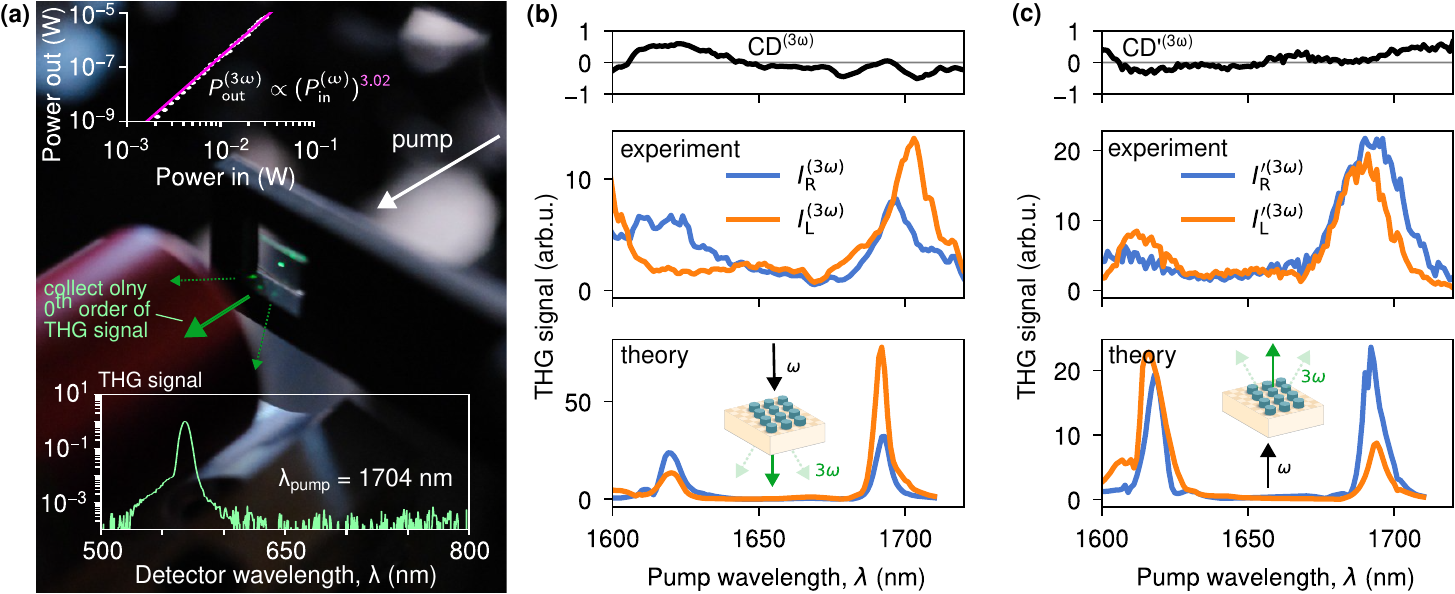}
    \caption{
    \textbf{Nonlinear experiment}.
    \textbf{(a)} Photo of the experimental setup. The intensity of the THG is sufficient to observe it by the naked eye.
    Inset top: power to power dependence is well fitted by $P_{\text{out}}^{(3\omega)} \approx 0.27 (P_{\text{in}}^{(\omega)} )^{3.02}$ which is a sign that undepleted pump approximation is valid.
    Inset bottom: typical THG signal spectrum at the detector.
    \textbf{(b)},\textbf{(c)} Total THG signal for RCP and LCP inputs and calculated nonlinear circular dichroism for excitations from the air and from the substrate. One can see a significant difference indicating that the fabricated metasurface indeed exhibits strong intrinsic nonlinear chirality. Theoretical spectra are shifted by the $+60$~nm to match the positions of peaks observed experimentally.
    }
    \label{fig:nonlinear_exp}
\end{figure*}

\textit{Nonlinear circular dichroism.}---While linear circular dichroism is well defined via the partial transmission, the definition of the \textit{nonlinear circular dichroism}, $\mathrm{CD}^{(n \omega)}$, is not fully settled in the literature.
The common option is to define nonlinear CD via total transmitted intensities from RCP and LCP inputs~\cite{SM,Petralli-Mallow1993JPhysChem,Kim2020NanoLett,Frizyuk2021NanoLett,Tang2020LaserPhotonicsRev}
\begin{equation}
    \mathrm{CD}^{(3\omega)} = \frac{I_{\text{R}}^{(3\omega)} - I_{\text{L}}^{(3\omega)}}{I_{\text{R}}^{(3\omega)} + I_{\text{L}}^{(3\omega)}} 
    \label{eq:NCDtot}
\end{equation}
where $I^{(3\omega)}_{\text{R(L)}}$ is the intensity of the third harmonic generation (THG) signal for RCP (LCP) input beam on the fundamental frequency.
Surprisingly, using nonlinear coupled mode theory it is possible to derive a very compact expression for the nonlinear $\mathrm{CD}^{(3\omega)}$ via coupling parameters given by Eq.~\eqref{eq:m} once there is a prominent resonance at the pump frequency, i.e. $\omega \approx \Re ( \omega_n)$, so \(\mathrm{CD}^{(3\omega)} \approx \mathrm{CD}^{(3\omega)}_n\), where~\cite{SM,Koshelev2023ACSPhot}
\begin{equation}
    \mathrm{CD}^{(3\omega)}_n = \frac{|m_{n\text{R}}|^6-|m_{n\text{L}}|^6}{|m_{n\text{R}}|^6+|m_{n\text{L}}|^6}.
    \label{eq:NCDn}
\end{equation}
For the excitation from the substrate one must substitute primed $m$-coupling parameters into Eq.~\eqref{eq:NCDn}. In the absence of top-down mirror symmetry, in general, $|m_{n\text{R(L)}}| \neq |m^{\prime}_{n\text{R(L)}}|$. 
Only recently the relation between the symmetry constraints and output signals in nonlinear transmission was studied~\cite{Koshelev2024JOPT}. In particular, in the presence of in-plane mirror symmetries $\mathrm{CD}^{(3\omega)}$ must be zero. For the monoclinic lattice, $\mathrm{CD}^{(3\omega)} \neq 0$ is allowed.
Remarkably, the true geometric chirality is not required for this, as even in the presence of out-of-plane mirror symmetry plane, different third harmonic signals can be generated by RCP or LCP pumping waves. {We note that this effect was previously associated with asymmetric nonlinear transmission in metallic metasurfaces~\cite{Achouri2018IEEE}, and can be predicted from the symmetry properties of full nonlinear scattering matrix}.  
In such case, however, the sign of nonlinear CD \eqref{eq:NCDtot} has to be opposite for the pumping incident on different metasurface sides~\cite{SM,Koshelev2024JOPT}.
In the case studied here, the geometric chirality is ensured by the presence of a transparent substrate. See Table~\ref{tab:summary} for the summary of the linear and nonlinear chiral characteristics.

In contrast to linear mode CD \eqref{eq:CDmode}, which is constructed from the symmetric combination of primed and nonprimed coupling parameters \eqref{eq:m}, the mode nonlinear CD \eqref{eq:NCDn} is \textit{side dependent}.
Structures with out-of-plane mirror symmetry must have $\mathrm{CD}_n^{(3\omega)} = - \mathrm{CD}_n^{\prime (3\omega)}$, where prime means that one should use $m^{\prime}_{n\text{R},\text{L}}$ in Eq.~\eqref{eq:NCDtot}. For our design we have the most general case with $|\mathrm{CD}_n^{(3\omega)}| \neq |\mathrm{CD}_n^{\prime (3\omega)}|$.

Since THG is a low efficient process, its \textit{resonant} enhancement is crucial for any potential practical application.
We employ resonant at the fundamental harmonic expecting signal enhancement by factor of $Q^3$, which is the highest possible excluding double resonant conditions requiring sophisticated designs~\cite{Koshelev2020Science}. The effect of the chiral resonance near THG frequency is discussed in Ref.~\cite{Antonov2023ieee_proceeding}.

We study the nonlinear process numerically by varying the lattice angle in the same manner as in Fig.~\ref{fig:concept}{(b)} and examine the THG signal in transmission collecting only the $0$th diffraction order. 
Nonlinear simulations predicts that for a chiral mode, the asymmetry in coupling strength  with different light polarizations ($\abs{m_{n\text{R}}}$ or $\abs{m_{n\text{L}}}$) manifests itself in the nonlinear signal as well. 
This leads to a stronger interaction with the preferred polarization, resulting in a higher intensity of the third-harmonic generation signal. See~\cite{SM} for more details.

We also prove this prediction experimentally, Fig.~\ref{fig:nonlinear_exp}. To study nonlinear properties of metasurfaces we use a tunable near-IR laser with a pulse duration of 500 fs and repetition rate of 5.14 MHz. Figure~\ref{fig:nonlinear_exp} presents the observed THG signal pumped with a peak laser intensity of 4~GW/cm$^2$: strong enough to be visible to the naked eye. The THG spectrum exited at the resonant wavelength of 1704 nm measured in transmission is shown in Figs.~\ref{fig:nonlinear_exp}(b),\ref{fig:nonlinear_exp}(c). 
The THG spectrum has a maximum at wavelength 568 nm and linewidth of 7 nm. 
We also record the power dependence of THG shown in the inset of Fig.~\ref{fig:nonlinear_exp}(a). It is well approximated by a cubic law as $P_{\text{out}}^{(3\omega)} \approx 0.27 (P_{\text{in}}^{(\omega)} )^{3.02}$ which is a sign that undepleted pump approximation is valid. The THG efficiency of 2$\times 10^{-4}$ is observed at the resonant pump with power of $30$~mW.

To investigate the resonant properties of THG generated from the metasurface, we pump it by a femtosecond laser with wavelengths in the range of 1500--1730 nm with a 1~nm step from two sides: from the air and from the substrate. The recorded THG spectra for RCP and LCP excitation show two distinct peaks corresponding to the two chiral modes studied above, see them compared with the theoretical THG spectra in Figs.~\ref{fig:nonlinear_exp}(b),\ref{fig:nonlinear_exp}(c). In all cases, THG signals demonstrate resonant enhancement near two wavelengths corresponding to the resonant modes observed in the linear regime. Next, we evaluate the experimental nonlinear CD according to Eq.~(\ref{eq:NCDtot}) as seen on the top panels of Figs.~\ref{fig:nonlinear_exp}(b),\ref{fig:nonlinear_exp}(c). Being driven by the resonances, nonlinear CD varies from $-0.64$ to $0.32$ for different pump wavelengths.

\textit{Conclusions.---}We have proposed and verified that the monoclinic geometry of metasurface lattices leads to distinct intrinsic chiral optical response, both in the linear and nonlinear regimes, even if metasurfaces are composed of highly symmetric achiral elements. We have developed a general analytical approach to characterize such metasurfaces by the mode chirality parameter as a quantitative measure of the eigenmode contribution to linear and nonlinear optical chirality. We have demonstrated experimentally the intrinsic chiral response of monoclinic silicon metasurfaces in both linear and nonlinear regimes.

\begin{acknowledgements}
Y.K. thanks Kuniaki Konishi, Olivier Martin, Michael Scalora, and Yuri Svirko for useful discussions.  The authors thank Shumin Xiao for valuable suggestions on the fabrication. This work was supported by the Australian Research Council (Grant No. DP210101292), the International Technology Center Indo-Pacific (ITC IPAC) via Army Research Office (contract FA520923C0023), National Natural Science Foundation of China (Grants 12261131500 and 12025402), and Shenzhen Fundamental Research Projects (grant JCYJ20210324120402006). M.G. acknowledges a support from the Russian Science Foundation (Project 23-42-00091).
\end{acknowledgements}

\bibliography{refs}

\begin{thebibliography}{50}%
\makeatletter
\providecommand \@ifxundefined [1]{%
 \@ifx{#1\undefined}
}%
\providecommand \@ifnum [1]{%
 \ifnum #1\expandafter \@firstoftwo
 \else \expandafter \@secondoftwo
 \fi
}%
\providecommand \@ifx [1]{%
 \ifx #1\expandafter \@firstoftwo
 \else \expandafter \@secondoftwo
 \fi
}%
\providecommand \natexlab [1]{#1}%
\providecommand \enquote  [1]{``#1''}%
\providecommand \bibnamefont  [1]{#1}%
\providecommand \bibfnamefont [1]{#1}%
\providecommand \citenamefont [1]{#1}%
\providecommand \href@noop [0]{\@secondoftwo}%
\providecommand \href [0]{\begingroup \@sanitize@url \@href}%
\providecommand \@href[1]{\@@startlink{#1}\@@href}%
\providecommand \@@href[1]{\endgroup#1\@@endlink}%
\providecommand \@sanitize@url [0]{\catcode `\\12\catcode `\$12\catcode
  `\&12\catcode `\#12\catcode `\^12\catcode `\_12\catcode `\%12\relax}%
\providecommand \@@startlink[1]{}%
\providecommand \@@endlink[0]{}%
\providecommand \url  [0]{\begingroup\@sanitize@url \@url }%
\providecommand \@url [1]{\endgroup\@href {#1}{\urlprefix }}%
\providecommand \urlprefix  [0]{URL }%
\providecommand \Eprint [0]{\href }%
\providecommand \doibase [0]{https://doi.org/}%
\providecommand \selectlanguage [0]{\@gobble}%
\providecommand \bibinfo  [0]{\@secondoftwo}%
\providecommand \bibfield  [0]{\@secondoftwo}%
\providecommand \translation [1]{[#1]}%
\providecommand \BibitemOpen [0]{}%
\providecommand \bibitemStop [0]{}%
\providecommand \bibitemNoStop [0]{.\EOS\space}%
\providecommand \EOS [0]{\spacefactor3000\relax}%
\providecommand \BibitemShut  [1]{\csname bibitem#1\endcsname}%
\let\auto@bib@innerbib\@empty
\bibitem [{\citenamefont {Khaliq}\ \emph {et~al.}(2023)\citenamefont {Khaliq},
  \citenamefont {Nauman}, \citenamefont {Lee},\ and\ \citenamefont
  {Kim}}]{Khaliq2023AOM}%
  \BibitemOpen
  \bibfield  {author} {\bibinfo {author} {\bibfnamefont {H.~S.}\ \bibnamefont
  {Khaliq}}, \bibinfo {author} {\bibfnamefont {A.}~\bibnamefont {Nauman}},
  \bibinfo {author} {\bibfnamefont {J.-W.}\ \bibnamefont {Lee}},\ and\ \bibinfo
  {author} {\bibfnamefont {H.-R.}\ \bibnamefont {Kim}},\ }\bibfield  {title}
  {\bibinfo {title} {{Recent Progress on Plasmonic and Dielectric Chiral
  Metasurfaces: Fundamentals, Design Strategies, and Implementation}},\ }\href
  {https://doi.org/10.1002/adom.202300644} {\bibfield  {journal} {\bibinfo
  {journal} {Adv. Opt. Mater.}\ }\textbf {\bibinfo {volume} {11}},\ \bibinfo
  {pages} {2300644} (\bibinfo {year} {2023})}\BibitemShut {NoStop}%
\bibitem [{\citenamefont {{\ifmmode\acute{A}\else\'{A}\fi}valos Ovando}\ \emph
  {et~al.}(2022)\citenamefont {{\ifmmode\acute{A}\else\'{A}\fi}valos Ovando},
  \citenamefont {Santiago}, \citenamefont {Movsesyan}, \citenamefont {Kong},
  \citenamefont {Yu}, \citenamefont {Besteiro}, \citenamefont {Khorashad},
  \citenamefont {Okamoto}, \citenamefont {Slocik}, \citenamefont
  {Correa-Duarte}, \citenamefont
  {Comesa{\ifmmode\tilde{n}\else\~{n}\fi}a-Hermo}, \citenamefont {Liedl},
  \citenamefont {Wang}, \citenamefont {Markovich}, \citenamefont {Burger},\
  and\ \citenamefont {Govorov}}]{Avalos-Ovando2022Jul}%
  \BibitemOpen
  \bibfield  {author} {\bibinfo {author} {\bibfnamefont {O.}~\bibnamefont
  {{\ifmmode\acute{A}\else\'{A}\fi}valos Ovando}}, \bibinfo {author}
  {\bibfnamefont {E.~Y.}\ \bibnamefont {Santiago}}, \bibinfo {author}
  {\bibfnamefont {A.}~\bibnamefont {Movsesyan}}, \bibinfo {author}
  {\bibfnamefont {X.-T.}\ \bibnamefont {Kong}}, \bibinfo {author}
  {\bibfnamefont {P.}~\bibnamefont {Yu}}, \bibinfo {author} {\bibfnamefont
  {L.~V.}\ \bibnamefont {Besteiro}}, \bibinfo {author} {\bibfnamefont {L.~K.}\
  \bibnamefont {Khorashad}}, \bibinfo {author} {\bibfnamefont {H.}~\bibnamefont
  {Okamoto}}, \bibinfo {author} {\bibfnamefont {J.~M.}\ \bibnamefont {Slocik}},
  \bibinfo {author} {\bibfnamefont {M.~A.}\ \bibnamefont {Correa-Duarte}},
  \bibinfo {author} {\bibfnamefont {M.}~\bibnamefont
  {Comesa{\ifmmode\tilde{n}\else\~{n}\fi}a-Hermo}}, \bibinfo {author}
  {\bibfnamefont {T.}~\bibnamefont {Liedl}}, \bibinfo {author} {\bibfnamefont
  {Z.}~\bibnamefont {Wang}}, \bibinfo {author} {\bibfnamefont {G.}~\bibnamefont
  {Markovich}}, \bibinfo {author} {\bibfnamefont {S.}~\bibnamefont {Burger}},\
  and\ \bibinfo {author} {\bibfnamefont {A.~O.}\ \bibnamefont {Govorov}},\
  }\bibfield  {title} {\bibinfo {title} {{Chiral Bioinspired Plasmonics: A
  Paradigm Shift for Optical Activity and Photochemistry}},\ }\href
  {https://doi.org/10.1021/acsphotonics.2c00445} {\bibfield  {journal}
  {\bibinfo  {journal} {ACS Photonics}\ }\textbf {\bibinfo {volume} {9}},\
  \bibinfo {pages} {2219} (\bibinfo {year} {2022})}\BibitemShut {NoStop}%
\bibitem [{\citenamefont {He}\ \emph {et~al.}(2018)\citenamefont {He},
  \citenamefont {Yang}, \citenamefont {Kuai}, \citenamefont {Shan},
  \citenamefont {Yang}, \citenamefont {Hu}, \citenamefont {Zhang},
  \citenamefont {Zhang},\ and\ \citenamefont {Zou}}]{He2018NC}%
  \BibitemOpen
  \bibfield  {author} {\bibinfo {author} {\bibfnamefont {C.}~\bibnamefont
  {He}}, \bibinfo {author} {\bibfnamefont {G.}~\bibnamefont {Yang}}, \bibinfo
  {author} {\bibfnamefont {Y.}~\bibnamefont {Kuai}}, \bibinfo {author}
  {\bibfnamefont {S.}~\bibnamefont {Shan}}, \bibinfo {author} {\bibfnamefont
  {L.}~\bibnamefont {Yang}}, \bibinfo {author} {\bibfnamefont {J.}~\bibnamefont
  {Hu}}, \bibinfo {author} {\bibfnamefont {D.}~\bibnamefont {Zhang}}, \bibinfo
  {author} {\bibfnamefont {Q.}~\bibnamefont {Zhang}},\ and\ \bibinfo {author}
  {\bibfnamefont {G.}~\bibnamefont {Zou}},\ }\bibfield  {title} {\bibinfo
  {title} {{Dissymmetry enhancement in enantioselective synthesis of helical
  polydiacetylene by application of superchiral light}},\ }\href
  {https://doi.org/10.1038/s41467-018-07533-y} {\bibfield  {journal} {\bibinfo
  {journal} {Nat. Commun.}\ }\textbf {\bibinfo {volume} {9}},\ \bibinfo {pages}
  {1} (\bibinfo {year} {2018})}\BibitemShut {NoStop}%
\bibitem [{\citenamefont {Lodahl}\ \emph {et~al.}(2017)\citenamefont {Lodahl},
  \citenamefont {Mahmoodian}, \citenamefont {Stobbe}, \citenamefont
  {Rauschenbeutel}, \citenamefont {Schneeweiss}, \citenamefont {Volz},
  \citenamefont {Pichler},\ and\ \citenamefont {Zoller}}]{Lodahl2017N}%
  \BibitemOpen
  \bibfield  {author} {\bibinfo {author} {\bibfnamefont {P.}~\bibnamefont
  {Lodahl}}, \bibinfo {author} {\bibfnamefont {S.}~\bibnamefont {Mahmoodian}},
  \bibinfo {author} {\bibfnamefont {S.}~\bibnamefont {Stobbe}}, \bibinfo
  {author} {\bibfnamefont {A.}~\bibnamefont {Rauschenbeutel}}, \bibinfo
  {author} {\bibfnamefont {P.}~\bibnamefont {Schneeweiss}}, \bibinfo {author}
  {\bibfnamefont {J.}~\bibnamefont {Volz}}, \bibinfo {author} {\bibfnamefont
  {H.}~\bibnamefont {Pichler}},\ and\ \bibinfo {author} {\bibfnamefont
  {P.}~\bibnamefont {Zoller}},\ }\bibfield  {title} {\bibinfo {title} {{Chiral
  quantum optics}},\ }\href {https://doi.org/10.1038/nature21037} {\bibfield
  {journal} {\bibinfo  {journal} {Nature}\ }\textbf {\bibinfo {volume} {541}},\
  \bibinfo {pages} {473} (\bibinfo {year} {2017})}\BibitemShut {NoStop}%
\bibitem [{\citenamefont {Papakostas}\ \emph {et~al.}(2003)\citenamefont
  {Papakostas}, \citenamefont {Potts}, \citenamefont {Bagnall}, \citenamefont
  {Prosvirnin}, \citenamefont {Coles},\ and\ \citenamefont
  {Zheludev}}]{Papakostas2003Mar}%
  \BibitemOpen
  \bibfield  {author} {\bibinfo {author} {\bibfnamefont {A.}~\bibnamefont
  {Papakostas}}, \bibinfo {author} {\bibfnamefont {A.}~\bibnamefont {Potts}},
  \bibinfo {author} {\bibfnamefont {D.~M.}\ \bibnamefont {Bagnall}}, \bibinfo
  {author} {\bibfnamefont {S.~L.}\ \bibnamefont {Prosvirnin}}, \bibinfo
  {author} {\bibfnamefont {H.~J.}\ \bibnamefont {Coles}},\ and\ \bibinfo
  {author} {\bibfnamefont {N.~I.}\ \bibnamefont {Zheludev}},\ }\bibfield
  {title} {\bibinfo {title} {{Optical Manifestations of Planar Chirality}},\
  }\href {https://doi.org/10.1103/PhysRevLett.90.107404} {\bibfield  {journal}
  {\bibinfo  {journal} {Phys. Rev. Lett.}\ }\textbf {\bibinfo {volume} {90}},\
  \bibinfo {pages} {107404} (\bibinfo {year} {2003})}\BibitemShut {NoStop}%
\bibitem [{\citenamefont {Ren}\ \emph {et~al.}(2012)\citenamefont {Ren},
  \citenamefont {Plum}, \citenamefont {Xu},\ and\ \citenamefont
  {Zheludev}}]{Ren2012May}%
  \BibitemOpen
  \bibfield  {author} {\bibinfo {author} {\bibfnamefont {M.}~\bibnamefont
  {Ren}}, \bibinfo {author} {\bibfnamefont {E.}~\bibnamefont {Plum}}, \bibinfo
  {author} {\bibfnamefont {J.}~\bibnamefont {Xu}},\ and\ \bibinfo {author}
  {\bibfnamefont {N.~I.}\ \bibnamefont {Zheludev}},\ }\bibfield  {title}
  {\bibinfo {title} {{Giant nonlinear optical activity in a plasmonic
  metamaterial}},\ }\href {https://doi.org/10.1038/ncomms1805} {\bibfield
  {journal} {\bibinfo  {journal} {Nat. Commun.}\ }\textbf {\bibinfo {volume}
  {3}},\ \bibinfo {pages} {1} (\bibinfo {year} {2012})}\BibitemShut {NoStop}%
\bibitem [{\citenamefont {Tanaka}\ \emph {et~al.}(2020)\citenamefont {Tanaka},
  \citenamefont {Arslan}, \citenamefont {Fasold}, \citenamefont {Steinert},
  \citenamefont {Sautter}, \citenamefont {Falkner}, \citenamefont {Pertsch},
  \citenamefont {Decker},\ and\ \citenamefont {Staude}}]{Tanaka2020Nov}%
  \BibitemOpen
  \bibfield  {author} {\bibinfo {author} {\bibfnamefont {K.}~\bibnamefont
  {Tanaka}}, \bibinfo {author} {\bibfnamefont {D.}~\bibnamefont {Arslan}},
  \bibinfo {author} {\bibfnamefont {S.}~\bibnamefont {Fasold}}, \bibinfo
  {author} {\bibfnamefont {M.}~\bibnamefont {Steinert}}, \bibinfo {author}
  {\bibfnamefont {J.}~\bibnamefont {Sautter}}, \bibinfo {author} {\bibfnamefont
  {M.}~\bibnamefont {Falkner}}, \bibinfo {author} {\bibfnamefont
  {T.}~\bibnamefont {Pertsch}}, \bibinfo {author} {\bibfnamefont
  {M.}~\bibnamefont {Decker}},\ and\ \bibinfo {author} {\bibfnamefont
  {I.}~\bibnamefont {Staude}},\ }\bibfield  {title} {\bibinfo {title} {{Chiral
  Bilayer All-Dielectric Metasurfaces}},\ }\href
  {https://doi.org/10.1021/acsnano.0c07295} {\bibfield  {journal} {\bibinfo
  {journal} {ACS Nano}\ }\textbf {\bibinfo {volume} {14}},\ \bibinfo {pages}
  {15926} (\bibinfo {year} {2020})}\BibitemShut {NoStop}%
\bibitem [{\citenamefont {Koshelev}\ \emph {et~al.}(2023)\citenamefont
  {Koshelev}, \citenamefont {Tang}, \citenamefont {Hu}, \citenamefont
  {Kravchenko}, \citenamefont {Li},\ and\ \citenamefont
  {Kivshar}}]{Koshelev2023ACSPhot}%
  \BibitemOpen
  \bibfield  {author} {\bibinfo {author} {\bibfnamefont {K.}~\bibnamefont
  {Koshelev}}, \bibinfo {author} {\bibfnamefont {Y.}~\bibnamefont {Tang}},
  \bibinfo {author} {\bibfnamefont {Z.}~\bibnamefont {Hu}}, \bibinfo {author}
  {\bibfnamefont {I.~I.}\ \bibnamefont {Kravchenko}}, \bibinfo {author}
  {\bibfnamefont {G.}~\bibnamefont {Li}},\ and\ \bibinfo {author}
  {\bibfnamefont {Y.}~\bibnamefont {Kivshar}},\ }\bibfield  {title} {\bibinfo
  {title} {{Resonant Chiral Effects in Nonlinear Dielectric Metasurfaces}},\
  }\href {https://doi.org/10.1021/acsphotonics.2c01926} {\bibfield  {journal}
  {\bibinfo  {journal} {ACS Photonics}\ }\textbf {\bibinfo {volume} {10}},\
  \bibinfo {pages} {298} (\bibinfo {year} {2023})}\BibitemShut {NoStop}%
\bibitem [{\citenamefont {Movsesyan}\ \emph {et~al.}(2022)\citenamefont
  {Movsesyan}, \citenamefont {Besteiro}, \citenamefont {Kong}, \citenamefont
  {Wang},\ and\ \citenamefont {Govorov}}]{Movsesyan2022Jul}%
  \BibitemOpen
  \bibfield  {author} {\bibinfo {author} {\bibfnamefont {A.}~\bibnamefont
  {Movsesyan}}, \bibinfo {author} {\bibfnamefont {L.~V.}\ \bibnamefont
  {Besteiro}}, \bibinfo {author} {\bibfnamefont {X.-T.}\ \bibnamefont {Kong}},
  \bibinfo {author} {\bibfnamefont {Z.}~\bibnamefont {Wang}},\ and\ \bibinfo
  {author} {\bibfnamefont {A.~O.}\ \bibnamefont {Govorov}},\ }\bibfield
  {title} {\bibinfo {title} {{Engineering Strongly Chiral Plasmonic Lattices
  with Achiral Unit Cells for Sensing and Photodetection}},\ }\href
  {https://doi.org/10.1002/adom.202101943} {\bibfield  {journal} {\bibinfo
  {journal} {Adv. Opt. Mater.}\ }\textbf {\bibinfo {volume} {10}},\ \bibinfo
  {pages} {2101943} (\bibinfo {year} {2022})}\BibitemShut {NoStop}%
\bibitem [{\citenamefont {Gryb}\ \emph {et~al.}(2023)\citenamefont {Gryb},
  \citenamefont {Wendisch}, \citenamefont {Aigner}, \citenamefont
  {G{\ifmmode\ddot{o}\else\"{o}\fi}lz}, \citenamefont {Tittl}, \citenamefont
  {de~S.~Menezes},\ and\ \citenamefont {Maier}}]{Gryb2023Oct}%
  \BibitemOpen
  \bibfield  {author} {\bibinfo {author} {\bibfnamefont {D.}~\bibnamefont
  {Gryb}}, \bibinfo {author} {\bibfnamefont {F.~J.}\ \bibnamefont {Wendisch}},
  \bibinfo {author} {\bibfnamefont {A.}~\bibnamefont {Aigner}}, \bibinfo
  {author} {\bibfnamefont {T.}~\bibnamefont
  {G{\ifmmode\ddot{o}\else\"{o}\fi}lz}}, \bibinfo {author} {\bibfnamefont
  {A.}~\bibnamefont {Tittl}}, \bibinfo {author} {\bibfnamefont
  {L.}~\bibnamefont {de~S.~Menezes}},\ and\ \bibinfo {author} {\bibfnamefont
  {S.~A.}\ \bibnamefont {Maier}},\ }\bibfield  {title} {\bibinfo {title}
  {{Two-Dimensional Chiral Metasurfaces Obtained by Geometrically Simple
  Meta-atom Rotations}},\ }\href {https://doi.org/10.1021/acs.nanolett.3c02168}
  {\bibfield  {journal} {\bibinfo  {journal} {Nano Lett.}\ }\textbf {\bibinfo
  {volume} {23}},\ \bibinfo {pages} {8891} (\bibinfo {year}
  {2023})}\BibitemShut {NoStop}%
\bibitem [{\citenamefont {Lyu}\ \emph {et~al.}(2023)\citenamefont {Lyu},
  \citenamefont {Li}, \citenamefont {Jia}, \citenamefont {Li}, \citenamefont
  {Yang}, \citenamefont {Cao}, \citenamefont {Kou}, \citenamefont {Liu},
  \citenamefont {Cao}, \citenamefont {Li},\ and\ \citenamefont
  {Shi}}]{Lyu2023LPR}%
  \BibitemOpen
  \bibfield  {author} {\bibinfo {author} {\bibfnamefont {B.}~\bibnamefont
  {Lyu}}, \bibinfo {author} {\bibfnamefont {Y.}~\bibnamefont {Li}}, \bibinfo
  {author} {\bibfnamefont {Q.}~\bibnamefont {Jia}}, \bibinfo {author}
  {\bibfnamefont {H.}~\bibnamefont {Li}}, \bibinfo {author} {\bibfnamefont
  {G.}~\bibnamefont {Yang}}, \bibinfo {author} {\bibfnamefont {F.}~\bibnamefont
  {Cao}}, \bibinfo {author} {\bibfnamefont {S.}~\bibnamefont {Kou}}, \bibinfo
  {author} {\bibfnamefont {D.}~\bibnamefont {Liu}}, \bibinfo {author}
  {\bibfnamefont {T.}~\bibnamefont {Cao}}, \bibinfo {author} {\bibfnamefont
  {G.}~\bibnamefont {Li}},\ and\ \bibinfo {author} {\bibfnamefont
  {J.}~\bibnamefont {Shi}},\ }\bibfield  {title} {\bibinfo {title}
  {{Manipulating the Chirality of Moir{\ifmmode\acute{e}\else\'{e}\fi}
  Metasurface by Symmetry Breaking}},\ }\href
  {https://doi.org/10.1002/lpor.202201004} {\bibfield  {journal} {\bibinfo
  {journal} {Laser Photonics Rev.}\ }\textbf {\bibinfo {volume} {17}},\
  \bibinfo {pages} {2201004} (\bibinfo {year} {2023})}\BibitemShut {NoStop}%
\bibitem [{\citenamefont {Wu}\ \emph {et~al.}(2018)\citenamefont {Wu},
  \citenamefont {Liu}, \citenamefont {Hill},\ and\ \citenamefont
  {Zheng}}]{Wu2018Nanoscale}%
  \BibitemOpen
  \bibfield  {author} {\bibinfo {author} {\bibfnamefont {Z.}~\bibnamefont
  {Wu}}, \bibinfo {author} {\bibfnamefont {Y.}~\bibnamefont {Liu}}, \bibinfo
  {author} {\bibfnamefont {E.~H.}\ \bibnamefont {Hill}},\ and\ \bibinfo
  {author} {\bibfnamefont {Y.}~\bibnamefont {Zheng}},\ }\bibfield  {title}
  {\bibinfo {title} {{Chiral metamaterials via
  Moir{\ifmmode\acute{e}\else\'{e}\fi} stacking}},\ }\href
  {https://doi.org/10.1039/C8NR04352C} {\bibfield  {journal} {\bibinfo
  {journal} {Nanoscale}\ }\textbf {\bibinfo {volume} {10}},\ \bibinfo {pages}
  {18096} (\bibinfo {year} {2018})}\BibitemShut {NoStop}%
\bibitem [{\citenamefont {Plum}\ \emph {et~al.}(2009)\citenamefont {Plum},
  \citenamefont {Fedotov},\ and\ \citenamefont {Zheludev}}]{Plum2009May}%
  \BibitemOpen
  \bibfield  {author} {\bibinfo {author} {\bibfnamefont {E.}~\bibnamefont
  {Plum}}, \bibinfo {author} {\bibfnamefont {V.~A.}\ \bibnamefont {Fedotov}},\
  and\ \bibinfo {author} {\bibfnamefont {N.~I.}\ \bibnamefont {Zheludev}},\
  }\bibfield  {title} {\bibinfo {title} {{Extrinsic electromagnetic chirality
  in metamaterials}},\ }\href {https://doi.org/10.1088/1464-4258/11/7/074009}
  {\bibfield  {journal} {\bibinfo  {journal} {J. Opt. A: Pure Appl. Opt.}\
  }\textbf {\bibinfo {volume} {11}},\ \bibinfo {pages} {074009} (\bibinfo
  {year} {2009})}\BibitemShut {NoStop}%
\bibitem [{\citenamefont {Asefa}\ \emph {et~al.}(2023)\citenamefont {Asefa},
  \citenamefont {Shim}, \citenamefont {Seong},\ and\ \citenamefont
  {Lee}}]{Asefa2023Sep}%
  \BibitemOpen
  \bibfield  {author} {\bibinfo {author} {\bibfnamefont {S.~A.}\ \bibnamefont
  {Asefa}}, \bibinfo {author} {\bibfnamefont {S.}~\bibnamefont {Shim}},
  \bibinfo {author} {\bibfnamefont {M.}~\bibnamefont {Seong}},\ and\ \bibinfo
  {author} {\bibfnamefont {D.}~\bibnamefont {Lee}},\ }\bibfield  {title}
  {\bibinfo {title} {{Chiral Metasurfaces: A Review of the Fundamentals and
  Research Advances}},\ }\href {https://doi.org/10.3390/app131910590}
  {\bibfield  {journal} {\bibinfo  {journal} {Appl. Sci.}\ }\textbf {\bibinfo
  {volume} {13}},\ \bibinfo {pages} {10590} (\bibinfo {year}
  {2023})}\BibitemShut {NoStop}%
\bibitem [{\citenamefont {Caloz}\ and\ \citenamefont
  {Sihvola}(2020)}]{Caloz2020Feb}%
  \BibitemOpen
  \bibfield  {author} {\bibinfo {author} {\bibfnamefont {C.}~\bibnamefont
  {Caloz}}\ and\ \bibinfo {author} {\bibfnamefont {A.}~\bibnamefont
  {Sihvola}},\ }\bibfield  {title} {\bibinfo {title} {{Electromagnetic
  Chirality, Part 1: The Microscopic Perspective [Electromagnetic
  Perspectives]}},\ }\href {https://doi.org/10.1109/MAP.2019.2955698}
  {\bibfield  {journal} {\bibinfo  {journal} {IEEE Antennas Propag. Mag.}\
  }\textbf {\bibinfo {volume} {62}},\ \bibinfo {pages} {58} (\bibinfo {year}
  {2020})}\BibitemShut {NoStop}%
\bibitem [{\citenamefont {Fasman}(2013)}]{fasman2013circular}%
  \BibitemOpen
  \bibfield  {author} {\bibinfo {author} {\bibfnamefont {G.~D.}\ \bibnamefont
  {Fasman}},\ }\href@noop {} {\emph {\bibinfo {title} {Circular dichroism and
  the conformational analysis of biomolecules}}}\ (\bibinfo  {publisher}
  {Springer Science \& Business Media},\ \bibinfo {year} {2013})\BibitemShut
  {NoStop}%
\bibitem [{\citenamefont {Kobayashi}\ and\ \citenamefont
  {Muranaka}(2011)}]{Kobayashi2011Nov}%
  \BibitemOpen
  \bibfield  {author} {\bibinfo {author} {\bibfnamefont {N.}~\bibnamefont
  {Kobayashi}}\ and\ \bibinfo {author} {\bibfnamefont {A.}~\bibnamefont
  {Muranaka}},\ }\href
  {https://books.google.com.au/books?id=mHIoDwAAQBAJ&dq=circular+dichroism+spectrograph+chemistry&lr=&source=gbs_navlinks_s}
  {\emph {\bibinfo {title} {{Circular Dichroism and Magnetic Circular Dichroism
  Spectroscopy for Organic Chemists}}}}\ (\bibinfo  {publisher} {Royal Society
  of Chemistry},\ \bibinfo {year} {2011})\BibitemShut {NoStop}%
\bibitem [{\citenamefont {Baase}\ and\ \citenamefont
  {Johnson}(1979)}]{Baase1979Feb}%
  \BibitemOpen
  \bibfield  {author} {\bibinfo {author} {\bibfnamefont {W.~A.}\ \bibnamefont
  {Baase}}\ and\ \bibinfo {author} {\bibfnamefont {W.~C.}\ \bibnamefont
  {Johnson}},\ }\bibfield  {title} {\bibinfo {title} {{Circular dichroism and
  DNA secondary structure}},\ }\href {https://doi.org/10.1093/nar/6.2.797}
  {\bibfield  {journal} {\bibinfo  {journal} {Nucleic Acids Res.}\ }\textbf
  {\bibinfo {volume} {6}},\ \bibinfo {pages} {797} (\bibinfo {year}
  {1979})}\BibitemShut {NoStop}%
\bibitem [{\citenamefont {Hu}\ \emph {et~al.}(2017)\citenamefont {Hu},
  \citenamefont {Zhao}, \citenamefont {Lin}, \citenamefont {Zhu}, \citenamefont
  {Zhu}, \citenamefont {Guo}, \citenamefont {Cao},\ and\ \citenamefont
  {Wang}}]{Hu2017Jan}%
  \BibitemOpen
  \bibfield  {author} {\bibinfo {author} {\bibfnamefont {J.}~\bibnamefont
  {Hu}}, \bibinfo {author} {\bibfnamefont {X.}~\bibnamefont {Zhao}}, \bibinfo
  {author} {\bibfnamefont {Y.}~\bibnamefont {Lin}}, \bibinfo {author}
  {\bibfnamefont {A.}~\bibnamefont {Zhu}}, \bibinfo {author} {\bibfnamefont
  {X.}~\bibnamefont {Zhu}}, \bibinfo {author} {\bibfnamefont {P.}~\bibnamefont
  {Guo}}, \bibinfo {author} {\bibfnamefont {B.}~\bibnamefont {Cao}},\ and\
  \bibinfo {author} {\bibfnamefont {C.}~\bibnamefont {Wang}},\ }\bibfield
  {title} {\bibinfo {title} {{All-dielectric metasurface circular dichroism
  waveplate}},\ }\href {https://doi.org/10.1038/srep41893} {\bibfield
  {journal} {\bibinfo  {journal} {Sci. Rep.}\ }\textbf {\bibinfo {volume}
  {7}},\ \bibinfo {pages} {1} (\bibinfo {year} {2017})}\BibitemShut {NoStop}%
\bibitem [{\citenamefont {Gorkunov}\ \emph {et~al.}(2020)\citenamefont
  {Gorkunov}, \citenamefont {Antonov},\ and\ \citenamefont
  {Kivshar}}]{Gorkunov2020Aug}%
  \BibitemOpen
  \bibfield  {author} {\bibinfo {author} {\bibfnamefont {M.~V.}\ \bibnamefont
  {Gorkunov}}, \bibinfo {author} {\bibfnamefont {A.~A.}\ \bibnamefont
  {Antonov}},\ and\ \bibinfo {author} {\bibfnamefont {Y.~S.}\ \bibnamefont
  {Kivshar}},\ }\bibfield  {title} {\bibinfo {title} {{Metasurfaces with
  Maximum Chirality Empowered by Bound States in the Continuum}},\ }\href
  {https://doi.org/10.1103/PhysRevLett.125.093903} {\bibfield  {journal}
  {\bibinfo  {journal} {Phys. Rev. Lett.}\ }\textbf {\bibinfo {volume} {125}},\
  \bibinfo {pages} {093903} (\bibinfo {year} {2020})}\BibitemShut {NoStop}%
\bibitem [{\citenamefont {Kim}\ and\ \citenamefont {Kim}(2020)}]{Kim2020Aug}%
  \BibitemOpen
  \bibfield  {author} {\bibinfo {author} {\bibfnamefont {K.-H.}\ \bibnamefont
  {Kim}}\ and\ \bibinfo {author} {\bibfnamefont {J.-R.}\ \bibnamefont {Kim}},\
  }\bibfield  {title} {\bibinfo {title} {{Dielectric Chiral Metasurfaces for
  Second-Harmonic Generation with Strong Circular Dichroism}},\ }\href
  {https://doi.org/10.1002/andp.202000078} {\bibfield  {journal} {\bibinfo
  {journal} {Ann. Phys.}\ }\textbf {\bibinfo {volume} {532}},\ \bibinfo {pages}
  {2000078} (\bibinfo {year} {2020})}\BibitemShut {NoStop}%
\bibitem [{\citenamefont {Koshelev}\ \emph {et~al.}(2024)\citenamefont
  {Koshelev}, \citenamefont {Toftul}, \citenamefont {Hwang},\ and\
  \citenamefont {Kivshar}}]{Koshelev2024JOPT}%
  \BibitemOpen
  \bibfield  {author} {\bibinfo {author} {\bibfnamefont {K.}~\bibnamefont
  {Koshelev}}, \bibinfo {author} {\bibfnamefont {I.}~\bibnamefont {Toftul}},
  \bibinfo {author} {\bibfnamefont {Y.}~\bibnamefont {Hwang}},\ and\ \bibinfo
  {author} {\bibfnamefont {Y.}~\bibnamefont {Kivshar}},\ }\bibfield  {title}
  {\bibinfo {title} {{Scattering matrix for chiral harmonic generation and
  frequency mixing in nonlinear metasurfaces}},\ }\href
  {https://doi.org/10.1088/2040-8986/ad3a78} {\bibfield  {journal} {\bibinfo
  {journal} {J. Opt.}\ }\textbf {\bibinfo {volume} {26}},\ \bibinfo {pages}
  {055003} (\bibinfo {year} {2024})}\BibitemShut {NoStop}%
\bibitem [{\citenamefont {Gorkunov}\ and\ \citenamefont
  {Antonov}(2024)}]{Gorkunov2024}%
  \BibitemOpen
  \bibfield  {author} {\bibinfo {author} {\bibfnamefont {M.~V.}\ \bibnamefont
  {Gorkunov}}\ and\ \bibinfo {author} {\bibfnamefont {A.~A.}\ \bibnamefont
  {Antonov}},\ }\bibinfo {title} {Rational design of maximum chiral dielectric
  metasurfaces},\ in\ \href
  {https://doi.org/10.1016/b978-0-32-395195-1.00014-4} {\emph {\bibinfo
  {booktitle} {All-Dielectric Nanophotonics}}}\ (\bibinfo  {publisher}
  {Elsevier},\ \bibinfo {year} {2024})\ pp.\ \bibinfo {pages}
  {243--286}\BibitemShut {NoStop}%
\bibitem [{\citenamefont {Overvig}\ \emph {et~al.}(2021)\citenamefont
  {Overvig}, \citenamefont {Yu},\ and\ \citenamefont {Alu}}]{Overvig2021Feb}%
  \BibitemOpen
  \bibfield  {author} {\bibinfo {author} {\bibfnamefont {A.}~\bibnamefont
  {Overvig}}, \bibinfo {author} {\bibfnamefont {N.}~\bibnamefont {Yu}},\ and\
  \bibinfo {author} {\bibfnamefont {A.}~\bibnamefont {Alu}},\ }\bibfield
  {title} {\bibinfo {title} {{Chiral Quasi-Bound States in the Continuum}},\
  }\href {https://doi.org/10.1103/PhysRevLett.126.073001} {\bibfield  {journal}
  {\bibinfo  {journal} {Phys. Rev. Lett.}\ }\textbf {\bibinfo {volume} {126}},\
  \bibinfo {pages} {073001} (\bibinfo {year} {2021})}\BibitemShut {NoStop}%
\bibitem [{\citenamefont {Kühner}\ \emph {et~al.}(2023)\citenamefont
  {Kühner}, \citenamefont {Wendisch}, \citenamefont {Antonov}, \citenamefont
  {Bürger}, \citenamefont {Hüttenhofer}, \citenamefont {De~S.~Menezes},
  \citenamefont {Maier}, \citenamefont {Gorkunov}, \citenamefont {Kivshar},\
  and\ \citenamefont {Tittl}}]{kuhner2023}%
  \BibitemOpen
  \bibfield  {author} {\bibinfo {author} {\bibfnamefont {L.}~\bibnamefont
  {Kühner}}, \bibinfo {author} {\bibfnamefont {F.~J.}\ \bibnamefont
  {Wendisch}}, \bibinfo {author} {\bibfnamefont {A.~A.}\ \bibnamefont
  {Antonov}}, \bibinfo {author} {\bibfnamefont {J.}~\bibnamefont {Bürger}},
  \bibinfo {author} {\bibfnamefont {L.}~\bibnamefont {Hüttenhofer}}, \bibinfo
  {author} {\bibfnamefont {L.}~\bibnamefont {De~S.~Menezes}}, \bibinfo {author}
  {\bibfnamefont {S.~A.}\ \bibnamefont {Maier}}, \bibinfo {author}
  {\bibfnamefont {M.~V.}\ \bibnamefont {Gorkunov}}, \bibinfo {author}
  {\bibfnamefont {Y.}~\bibnamefont {Kivshar}},\ and\ \bibinfo {author}
  {\bibfnamefont {A.}~\bibnamefont {Tittl}},\ }\bibfield  {title} {\bibinfo
  {title} {Unlocking the out-of-plane dimension for photonic bound states in
  the continuum to achieve maximum optical chirality},\ }\href
  {https://doi.org/10.1038/s41377-023-01295-z} {\bibfield  {journal} {\bibinfo
  {journal} {Light: Science \& Applications}\ }\textbf {\bibinfo {volume}
  {12}},\ \bibinfo {pages} {250} (\bibinfo {year} {2023})}\BibitemShut
  {NoStop}%
\bibitem [{SM()}]{SM}%
  \BibitemOpen
  \href@noop {} {\bibinfo {title} {{See Supplemental Material at \textit{[URL
  will be inserted by publisher]} for the numerical simulations methods;
  discussion of possible reasons of mismatch between modeling and experiment;
  numerical demonstration of the linear dependence of anti-crossing size as a
  function of $|a-b|$; background field formulation; linear and nonlinear
  chiral transmission characteristics using S-matrix theory; chiral coupled
  mode theory for the chiral transmission characteristics; and experimental
  methods. Supplemental Material includes
  Refs.~\cite{boyd2008nonlinear,CanosValero2024PhysRevRes,COMSOL_SHG_example,Fan2003JOSAA,Hermann1984,Igoshin2024PRB,koshelev2019nonlinear,kuhne2021fabrication,Maier2006OE,Malitson1965JOSA,Muljarov2018OL,Polyanskiy2024SciData,seok2011radiation,Shalin2024AllDielectricNanophotonics}
  }}}\BibitemShut {NoStop}%
\bibitem [{\citenamefont {Weiss}\ and\ \citenamefont
  {Muljarov}(2018)}]{Weiss2018PRB}%
  \BibitemOpen
  \bibfield  {author} {\bibinfo {author} {\bibfnamefont {T.}~\bibnamefont
  {Weiss}}\ and\ \bibinfo {author} {\bibfnamefont {E.~A.}\ \bibnamefont
  {Muljarov}},\ }\bibfield  {title} {\bibinfo {title} {{How to calculate the
  pole expansion of the optical scattering matrix from the resonant states}},\
  }\href {https://doi.org/10.1103/PhysRevB.98.085433} {\bibfield  {journal}
  {\bibinfo  {journal} {Phys. Rev. B}\ }\textbf {\bibinfo {volume} {98}},\
  \bibinfo {pages} {085433} (\bibinfo {year} {2018})}\BibitemShut {NoStop}%
\bibitem [{\citenamefont {Koshelev}(2022)}]{koshelev2022PhDthesis}%
  \BibitemOpen
  \bibfield  {author} {\bibinfo {author} {\bibfnamefont {K.}~\bibnamefont
  {Koshelev}},\ }\emph {\bibinfo {title} {Advanced trapping of light in
  resonant dielectric metastructures for nonlinear optics}},\ \href@noop {}
  {Ph.D. thesis},\ \bibinfo  {school} {The Australian National University
  (Australia)} (\bibinfo {year} {2022})\BibitemShut {NoStop}%
\bibitem [{\citenamefont {Koshelev}\ \emph {et~al.}(2020)\citenamefont
  {Koshelev}, \citenamefont {Kruk}, \citenamefont {Melik-Gaykazyan},
  \citenamefont {Choi}, \citenamefont {Bogdanov}, \citenamefont {Park},\ and\
  \citenamefont {Kivshar}}]{Koshelev2020Science}%
  \BibitemOpen
  \bibfield  {author} {\bibinfo {author} {\bibfnamefont {K.}~\bibnamefont
  {Koshelev}}, \bibinfo {author} {\bibfnamefont {S.}~\bibnamefont {Kruk}},
  \bibinfo {author} {\bibfnamefont {E.}~\bibnamefont {Melik-Gaykazyan}},
  \bibinfo {author} {\bibfnamefont {J.-H.}\ \bibnamefont {Choi}}, \bibinfo
  {author} {\bibfnamefont {A.}~\bibnamefont {Bogdanov}}, \bibinfo {author}
  {\bibfnamefont {H.-G.}\ \bibnamefont {Park}},\ and\ \bibinfo {author}
  {\bibfnamefont {Y.}~\bibnamefont {Kivshar}},\ }\bibfield  {title} {\bibinfo
  {title} {{Subwavelength dielectric resonators for nonlinear nanophotonics}},\
  }\href {https://doi.org/10.1126/science.aaz3985} {\bibfield  {journal}
  {\bibinfo  {journal} {Science}\ }\textbf {\bibinfo {volume} {367}},\ \bibinfo
  {pages} {288} (\bibinfo {year} {2020})}\BibitemShut {NoStop}%
\bibitem [{not()}]{note_one}%
  \BibitemOpen
  \href@noop {} {\bibinfo {title} {{It follows from the scattering formalism
  for resonant states~\cite{Weiss2018PRB} that one should substitute the
  complex eigenmode frequency $\omega_n$ into the background field
  }}}\BibitemShut {NoStop}%
\bibitem [{\citenamefont {Petralli-Mallow}\ \emph {et~al.}(1993)\citenamefont
  {Petralli-Mallow}, \citenamefont {Wong}, \citenamefont {Byers}, \citenamefont
  {Yee},\ and\ \citenamefont {Hicks}}]{Petralli-Mallow1993JPhysChem}%
  \BibitemOpen
  \bibfield  {author} {\bibinfo {author} {\bibfnamefont {T.}~\bibnamefont
  {Petralli-Mallow}}, \bibinfo {author} {\bibfnamefont {T.~M.}\ \bibnamefont
  {Wong}}, \bibinfo {author} {\bibfnamefont {J.~D.}\ \bibnamefont {Byers}},
  \bibinfo {author} {\bibfnamefont {H.~I.}\ \bibnamefont {Yee}},\ and\ \bibinfo
  {author} {\bibfnamefont {J.~M.}\ \bibnamefont {Hicks}},\ }\bibfield  {title}
  {\bibinfo {title} {{Circular dichroism spectroscopy at interfaces: a surface
  second harmonic generation study}},\ }\href
  {https://doi.org/10.1021/j100109a022} {\bibfield  {journal} {\bibinfo
  {journal} {J. Phys. Chem.}\ }\textbf {\bibinfo {volume} {97}},\ \bibinfo
  {pages} {1383} (\bibinfo {year} {1993})}\BibitemShut {NoStop}%
\bibitem [{\citenamefont {Kim}\ \emph {et~al.}(2020)\citenamefont {Kim},
  \citenamefont {Yu}, \citenamefont {Hwang}, \citenamefont {Park},
  \citenamefont {Demmerle}, \citenamefont {Boehm}, \citenamefont {Amann},
  \citenamefont {Belkin},\ and\ \citenamefont {Lee}}]{Kim2020NanoLett}%
  \BibitemOpen
  \bibfield  {author} {\bibinfo {author} {\bibfnamefont {D.}~\bibnamefont
  {Kim}}, \bibinfo {author} {\bibfnamefont {J.}~\bibnamefont {Yu}}, \bibinfo
  {author} {\bibfnamefont {I.}~\bibnamefont {Hwang}}, \bibinfo {author}
  {\bibfnamefont {S.}~\bibnamefont {Park}}, \bibinfo {author} {\bibfnamefont
  {F.}~\bibnamefont {Demmerle}}, \bibinfo {author} {\bibfnamefont
  {G.}~\bibnamefont {Boehm}}, \bibinfo {author} {\bibfnamefont {M.-C.}\
  \bibnamefont {Amann}}, \bibinfo {author} {\bibfnamefont {M.~A.}\ \bibnamefont
  {Belkin}},\ and\ \bibinfo {author} {\bibfnamefont {J.}~\bibnamefont {Lee}},\
  }\bibfield  {title} {\bibinfo {title} {{Giant Nonlinear Circular Dichroism
  from Intersubband Polaritonic Metasurfaces}},\ }\href
  {https://doi.org/10.1021/acs.nanolett.0c02978} {\bibfield  {journal}
  {\bibinfo  {journal} {Nano Lett.}\ }\textbf {\bibinfo {volume} {20}},\
  \bibinfo {pages} {8032} (\bibinfo {year} {2020})}\BibitemShut {NoStop}%
\bibitem [{\citenamefont {Frizyuk}\ \emph {et~al.}(2021)\citenamefont
  {Frizyuk}, \citenamefont {Melik-Gaykazyan}, \citenamefont {Choi},
  \citenamefont {Petrov}, \citenamefont {Park},\ and\ \citenamefont
  {Kivshar}}]{Frizyuk2021NanoLett}%
  \BibitemOpen
  \bibfield  {author} {\bibinfo {author} {\bibfnamefont {K.}~\bibnamefont
  {Frizyuk}}, \bibinfo {author} {\bibfnamefont {E.}~\bibnamefont
  {Melik-Gaykazyan}}, \bibinfo {author} {\bibfnamefont {J.-H.}\ \bibnamefont
  {Choi}}, \bibinfo {author} {\bibfnamefont {M.~I.}\ \bibnamefont {Petrov}},
  \bibinfo {author} {\bibfnamefont {H.-G.}\ \bibnamefont {Park}},\ and\
  \bibinfo {author} {\bibfnamefont {Y.}~\bibnamefont {Kivshar}},\ }\bibfield
  {title} {\bibinfo {title} {{Nonlinear Circular Dichroism in Mie-Resonant
  Nanoparticle Dimers}},\ }\href {https://doi.org/10.1021/acs.nanolett.1c01025}
  {\bibfield  {journal} {\bibinfo  {journal} {Nano Lett.}\ }\textbf {\bibinfo
  {volume} {21}},\ \bibinfo {pages} {4381} (\bibinfo {year}
  {2021})}\BibitemShut {NoStop}%
\bibitem [{\citenamefont {Tang}\ \emph {et~al.}(2020)\citenamefont {Tang},
  \citenamefont {Liu}, \citenamefont {Deng}, \citenamefont {Li}, \citenamefont
  {Li},\ and\ \citenamefont {Li}}]{Tang2020LaserPhotonicsRev}%
  \BibitemOpen
  \bibfield  {author} {\bibinfo {author} {\bibfnamefont {Y.}~\bibnamefont
  {Tang}}, \bibinfo {author} {\bibfnamefont {Z.}~\bibnamefont {Liu}}, \bibinfo
  {author} {\bibfnamefont {J.}~\bibnamefont {Deng}}, \bibinfo {author}
  {\bibfnamefont {K.}~\bibnamefont {Li}}, \bibinfo {author} {\bibfnamefont
  {J.}~\bibnamefont {Li}},\ and\ \bibinfo {author} {\bibfnamefont
  {G.}~\bibnamefont {Li}},\ }\bibfield  {title} {\bibinfo {title}
  {{Nano-Kirigami Metasurface with Giant Nonlinear Optical Circular
  Dichroism}},\ }\href {https://doi.org/10.1002/lpor.202000085} {\bibfield
  {journal} {\bibinfo  {journal} {Laser Photonics Rev.}\ }\textbf {\bibinfo
  {volume} {14}},\ \bibinfo {pages} {2000085} (\bibinfo {year}
  {2020})}\BibitemShut {NoStop}%
\bibitem [{\citenamefont {Achouri}\ \emph {et~al.}(2018)\citenamefont
  {Achouri}, \citenamefont {Bernasconi}, \citenamefont {Butet},\ and\
  \citenamefont {Martin}}]{Achouri2018IEEE}%
  \BibitemOpen
  \bibfield  {author} {\bibinfo {author} {\bibfnamefont {K.}~\bibnamefont
  {Achouri}}, \bibinfo {author} {\bibfnamefont {G.~D.}\ \bibnamefont
  {Bernasconi}}, \bibinfo {author} {\bibfnamefont {J.}~\bibnamefont {Butet}},\
  and\ \bibinfo {author} {\bibfnamefont {O.~J.~F.}\ \bibnamefont {Martin}},\
  }\bibfield  {title} {\bibinfo {title} {{Homogenization and Scattering
  Analysis of Second-Harmonic Generation in Nonlinear Metasurfaces}},\ }\href
  {https://doi.org/10.1109/TAP.2018.2863116} {\bibfield  {journal} {\bibinfo
  {journal} {IEEE Trans. Antennas Propag.}\ }\textbf {\bibinfo {volume} {66}},\
  \bibinfo {pages} {6061} (\bibinfo {year} {2018})}\BibitemShut {NoStop}%
\bibitem [{\citenamefont {Antonov}\ \emph {et~al.}()\citenamefont {Antonov},
  \citenamefont {Gorkunov},\ and\ \citenamefont
  {Kivshar}}]{Antonov2023ieee_proceeding}%
  \BibitemOpen
  \bibfield  {author} {\bibinfo {author} {\bibfnamefont {A.}~\bibnamefont
  {Antonov}}, \bibinfo {author} {\bibfnamefont {M.}~\bibnamefont {Gorkunov}},\
  and\ \bibinfo {author} {\bibfnamefont {Y.}~\bibnamefont {Kivshar}},\
  }\bibfield  {title} {\bibinfo {title} {{Chiral Harmonic Generation by
  Quasi-Bound States in the Continuum}},\ }in\ \href
  {https://doi.org/10.1109/Metamaterials58257.2023.10289562} {\emph {\bibinfo
  {booktitle} {{2023 Seventeenth International Congress on Artificial Materials
  for Novel Wave Phenomena (Metamaterials)}}}}\ (\bibinfo  {publisher} {IEEE})\
  pp.\ \bibinfo {pages} {11--16}\BibitemShut {NoStop}%
\bibitem [{\citenamefont {Boyd}\ \emph {et~al.}(2008)\citenamefont {Boyd},
  \citenamefont {Gaeta},\ and\ \citenamefont {Giese}}]{boyd2008nonlinear}%
  \BibitemOpen
  \bibfield  {author} {\bibinfo {author} {\bibfnamefont {R.~W.}\ \bibnamefont
  {Boyd}}, \bibinfo {author} {\bibfnamefont {A.~L.}\ \bibnamefont {Gaeta}},\
  and\ \bibinfo {author} {\bibfnamefont {E.}~\bibnamefont {Giese}},\
  }\href@noop {} {\emph {\bibinfo {title} {Springer Handbook of Atomic,
  Molecular, and Optical Physics}}}\ (\bibinfo  {publisher} {Springer},\
  \bibinfo {year} {2008})\ pp.\ \bibinfo {pages} {1097--1110}\BibitemShut
  {NoStop}%
\bibitem [{\citenamefont {Can{\ifmmode\acute{o}\else\'{o}\fi}s~Valero}\ \emph
  {et~al.}(2024)\citenamefont {Can{\ifmmode\acute{o}\else\'{o}\fi}s~Valero},
  \citenamefont {Bobrovs}, \citenamefont {Weiss}, \citenamefont {Gao},
  \citenamefont {Shalin},\ and\ \citenamefont
  {Kivshar}}]{CanosValero2024PhysRevRes}%
  \BibitemOpen
  \bibfield  {author} {\bibinfo {author} {\bibfnamefont {A.}~\bibnamefont
  {Can{\ifmmode\acute{o}\else\'{o}\fi}s~Valero}}, \bibinfo {author}
  {\bibfnamefont {V.}~\bibnamefont {Bobrovs}}, \bibinfo {author} {\bibfnamefont
  {T.}~\bibnamefont {Weiss}}, \bibinfo {author} {\bibfnamefont
  {L.}~\bibnamefont {Gao}}, \bibinfo {author} {\bibfnamefont {A.~S.}\
  \bibnamefont {Shalin}},\ and\ \bibinfo {author} {\bibfnamefont
  {Y.}~\bibnamefont {Kivshar}},\ }\bibfield  {title} {\bibinfo {title}
  {{Bianisotropic exceptional points in an isolated dielectric nanoparticle}},\
  }\href {https://doi.org/10.1103/PhysRevResearch.6.013053} {\bibfield
  {journal} {\bibinfo  {journal} {Phys. Rev. Res.}\ }\textbf {\bibinfo {volume}
  {6}},\ \bibinfo {pages} {013053} (\bibinfo {year} {2024})}\BibitemShut
  {NoStop}%
\bibitem [{COM(2024)}]{COMSOL_SHG_example}%
  \BibitemOpen
  \href
  {https://www.comsol.com/model/second-harmonic-generation-in-the-frequency-domain-24151}
  {\bibinfo {title} {{Second Harmonic Generation in the Frequency Domain}}}
  (\bibinfo {year} {2024}),\ \bibinfo {note} {[Online; accessed 22. Apr.
  2024]}\BibitemShut {NoStop}%
\bibitem [{\citenamefont {Fan}\ \emph {et~al.}(2003)\citenamefont {Fan},
  \citenamefont {Suh},\ and\ \citenamefont {Joannopoulos}}]{Fan2003JOSAA}%
  \BibitemOpen
  \bibfield  {author} {\bibinfo {author} {\bibfnamefont {S.}~\bibnamefont
  {Fan}}, \bibinfo {author} {\bibfnamefont {W.}~\bibnamefont {Suh}},\ and\
  \bibinfo {author} {\bibfnamefont {J.~D.}\ \bibnamefont {Joannopoulos}},\
  }\bibfield  {title} {\bibinfo {title} {{Temporal coupled-mode theory for the
  Fano resonance in optical resonators}},\ }\href
  {https://doi.org/10.1364/JOSAA.20.000569} {\bibfield  {journal} {\bibinfo
  {journal} {J. Opt. Soc. Am. A, JOSAA}\ }\textbf {\bibinfo {volume} {20}},\
  \bibinfo {pages} {569} (\bibinfo {year} {2003})}\BibitemShut {NoStop}%
\bibitem [{\citenamefont {Hermann}(1984)}]{Hermann1984}%
  \BibitemOpen
  \bibfield  {author} {\bibinfo {author} {\bibfnamefont {H.}~\bibnamefont
  {Hermann}},\ }\href@noop {} {\emph {\bibinfo {title} {Waves and {{Fields}} in
  {{Optoelectronics}}}}}\ (\bibinfo {year} {1984})\BibitemShut {NoStop}%
\bibitem [{\citenamefont {Igoshin}\ \emph {et~al.}(2024)\citenamefont
  {Igoshin}, \citenamefont {Tsimokha}, \citenamefont {Nikitina}, \citenamefont
  {Petrov}, \citenamefont {Toftul},\ and\ \citenamefont
  {Frizyuk}}]{Igoshin2024PRB}%
  \BibitemOpen
  \bibfield  {author} {\bibinfo {author} {\bibfnamefont {V.}~\bibnamefont
  {Igoshin}}, \bibinfo {author} {\bibfnamefont {M.}~\bibnamefont {Tsimokha}},
  \bibinfo {author} {\bibfnamefont {A.}~\bibnamefont {Nikitina}}, \bibinfo
  {author} {\bibfnamefont {M.}~\bibnamefont {Petrov}}, \bibinfo {author}
  {\bibfnamefont {I.}~\bibnamefont {Toftul}},\ and\ \bibinfo {author}
  {\bibfnamefont {K.}~\bibnamefont {Frizyuk}},\ }\bibfield  {title} {\bibinfo
  {title} {{Exceptional points in single open acoustic resonator due to
  symmetry breaking}},\ }\href {https://doi.org/10.1103/PhysRevB.109.144102}
  {\bibfield  {journal} {\bibinfo  {journal} {Phys. Rev. B}\ }\textbf {\bibinfo
  {volume} {109}},\ \bibinfo {pages} {144102} (\bibinfo {year}
  {2024})}\BibitemShut {NoStop}%
\bibitem [{\citenamefont {Koshelev}\ \emph {et~al.}(2019)\citenamefont
  {Koshelev}, \citenamefont {Tang}, \citenamefont {Li}, \citenamefont {Choi},
  \citenamefont {Li},\ and\ \citenamefont {Kivshar}}]{koshelev2019nonlinear}%
  \BibitemOpen
  \bibfield  {author} {\bibinfo {author} {\bibfnamefont {K.}~\bibnamefont
  {Koshelev}}, \bibinfo {author} {\bibfnamefont {Y.}~\bibnamefont {Tang}},
  \bibinfo {author} {\bibfnamefont {K.}~\bibnamefont {Li}}, \bibinfo {author}
  {\bibfnamefont {D.-Y.}\ \bibnamefont {Choi}}, \bibinfo {author}
  {\bibfnamefont {G.}~\bibnamefont {Li}},\ and\ \bibinfo {author}
  {\bibfnamefont {Y.}~\bibnamefont {Kivshar}},\ }\bibfield  {title} {\bibinfo
  {title} {Nonlinear metasurfaces governed by bound states in the continuum},\
  }\href@noop {} {\bibfield  {journal} {\bibinfo  {journal} {Acs Photonics}\
  }\textbf {\bibinfo {volume} {6}},\ \bibinfo {pages} {1639} (\bibinfo {year}
  {2019})}\BibitemShut {NoStop}%
\bibitem [{\citenamefont {K{\"u}hne}\ \emph {et~al.}(2021)\citenamefont
  {K{\"u}hne}, \citenamefont {Wang}, \citenamefont {Weber}, \citenamefont
  {K{\"u}hner}, \citenamefont {Maier},\ and\ \citenamefont
  {Tittl}}]{kuhne2021fabrication}%
  \BibitemOpen
  \bibfield  {author} {\bibinfo {author} {\bibfnamefont {J.}~\bibnamefont
  {K{\"u}hne}}, \bibinfo {author} {\bibfnamefont {J.}~\bibnamefont {Wang}},
  \bibinfo {author} {\bibfnamefont {T.}~\bibnamefont {Weber}}, \bibinfo
  {author} {\bibfnamefont {L.}~\bibnamefont {K{\"u}hner}}, \bibinfo {author}
  {\bibfnamefont {S.~A.}\ \bibnamefont {Maier}},\ and\ \bibinfo {author}
  {\bibfnamefont {A.}~\bibnamefont {Tittl}},\ }\bibfield  {title} {\bibinfo
  {title} {Fabrication robustness in bic metasurfaces},\ }\href@noop {}
  {\bibfield  {journal} {\bibinfo  {journal} {Nanophotonics}\ }\textbf
  {\bibinfo {volume} {10}},\ \bibinfo {pages} {4305} (\bibinfo {year}
  {2021})}\BibitemShut {NoStop}%
\bibitem [{\citenamefont {Maier}(2006)}]{Maier2006OE}%
  \BibitemOpen
  \bibfield  {author} {\bibinfo {author} {\bibfnamefont {S.~A.}\ \bibnamefont
  {Maier}},\ }\bibfield  {title} {\bibinfo {title} {Plasmonic field enhancement
  and {{SERS}} in the effective mode volume picture},\ }\href
  {https://doi.org/10.1364/OE.14.001957} {\bibfield  {journal} {\bibinfo
  {journal} {Opt. Express}\ }\textbf {\bibinfo {volume} {14}},\ \bibinfo
  {pages} {1957} (\bibinfo {year} {2006})}\BibitemShut {NoStop}%
\bibitem [{\citenamefont {Malitson}(1965)}]{Malitson1965JOSA}%
  \BibitemOpen
  \bibfield  {author} {\bibinfo {author} {\bibfnamefont {I.~H.}\ \bibnamefont
  {Malitson}},\ }\bibfield  {title} {\bibinfo {title} {Interspecimen comparison
  of the refractive index of fused silica},\ }\href
  {https://doi.org/10.1364/JOSA.55.001205} {\bibfield  {journal} {\bibinfo
  {journal} {JOSA}\ }\textbf {\bibinfo {volume} {55}},\ \bibinfo {pages} {1205}
  (\bibinfo {year} {1965})}\BibitemShut {NoStop}%
\bibitem [{\citenamefont {Muljarov}\ and\ \citenamefont
  {Weiss}(2018)}]{Muljarov2018OL}%
  \BibitemOpen
  \bibfield  {author} {\bibinfo {author} {\bibfnamefont {E.~A.}\ \bibnamefont
  {Muljarov}}\ and\ \bibinfo {author} {\bibfnamefont {T.}~\bibnamefont
  {Weiss}},\ }\bibfield  {title} {\bibinfo {title} {{Resonant-state expansion
  for open optical systems: generalization to magnetic, chiral, and
  bi-anisotropic materials}},\ }\href {https://doi.org/10.1364/OL.43.001978}
  {\bibfield  {journal} {\bibinfo  {journal} {Opt. Lett.}\ }\textbf {\bibinfo
  {volume} {43}},\ \bibinfo {pages} {1978} (\bibinfo {year}
  {2018})}\BibitemShut {NoStop}%
\bibitem [{\citenamefont {Polyanskiy}(2024)}]{Polyanskiy2024SciData}%
  \BibitemOpen
  \bibfield  {author} {\bibinfo {author} {\bibfnamefont {M.~N.}\ \bibnamefont
  {Polyanskiy}},\ }\bibfield  {title} {\bibinfo {title} {{Refractiveindex.info
  database of optical constants}},\ }\href
  {https://doi.org/10.1038/s41597-023-02898-2} {\bibfield  {journal} {\bibinfo
  {journal} {Sci. Data}\ }\textbf {\bibinfo {volume} {11}},\ \bibinfo {pages}
  {1} (\bibinfo {year} {2024})}\BibitemShut {NoStop}%
\bibitem [{\citenamefont {Seok}\ \emph {et~al.}(2011)\citenamefont {Seok},
  \citenamefont {Jamshidi}, \citenamefont {Kim}, \citenamefont {Dhuey},
  \citenamefont {Lakhani}, \citenamefont {Choo}, \citenamefont {Schuck},
  \citenamefont {Cabrini}, \citenamefont {Schwartzberg}, \citenamefont {Bokor}
  \emph {et~al.}}]{seok2011radiation}%
  \BibitemOpen
  \bibfield  {author} {\bibinfo {author} {\bibfnamefont {T.~J.}\ \bibnamefont
  {Seok}}, \bibinfo {author} {\bibfnamefont {A.}~\bibnamefont {Jamshidi}},
  \bibinfo {author} {\bibfnamefont {M.}~\bibnamefont {Kim}}, \bibinfo {author}
  {\bibfnamefont {S.}~\bibnamefont {Dhuey}}, \bibinfo {author} {\bibfnamefont
  {A.}~\bibnamefont {Lakhani}}, \bibinfo {author} {\bibfnamefont
  {H.}~\bibnamefont {Choo}}, \bibinfo {author} {\bibfnamefont {P.~J.}\
  \bibnamefont {Schuck}}, \bibinfo {author} {\bibfnamefont {S.}~\bibnamefont
  {Cabrini}}, \bibinfo {author} {\bibfnamefont {A.~M.}\ \bibnamefont
  {Schwartzberg}}, \bibinfo {author} {\bibfnamefont {J.}~\bibnamefont {Bokor}},
  \emph {et~al.},\ }\bibfield  {title} {\bibinfo {title} {Radiation engineering
  of optical antennas for maximum field enhancement},\ }\href@noop {}
  {\bibfield  {journal} {\bibinfo  {journal} {Nano letters}\ }\textbf {\bibinfo
  {volume} {11}},\ \bibinfo {pages} {2606} (\bibinfo {year}
  {2011})}\BibitemShut {NoStop}%
\bibitem [{\citenamefont {Shalin}\ \emph {et~al.}(2024)\citenamefont {Shalin},
  \citenamefont {Valero},\ and\ \citenamefont
  {Miroshnichenko}}]{Shalin2024AllDielectricNanophotonics}%
  \BibitemOpen
  \bibfield  {author} {\bibinfo {author} {\bibfnamefont {A.~S.}\ \bibnamefont
  {Shalin}}, \bibinfo {author} {\bibfnamefont {A.~C.}\ \bibnamefont {Valero}},\
  and\ \bibinfo {author} {\bibfnamefont {A.}~\bibnamefont {Miroshnichenko}},\
  }\href {https://doi.org/10.1016/C2021-0-02389-0} {\emph {\bibinfo {title}
  {{All-Dielectric Nanophotonics}}}}\ (\bibinfo  {publisher} {Elsevier},\
  \bibinfo {address} {Walthm, MA, USA},\ \bibinfo {year} {2024})\BibitemShut
  {NoStop}%
\end{thebibliography}%

\appendix
\newpage 
\onecolumngrid

\begin{center}
	\textbf{Supplementary Material} \\
	\textbf{Chiral dichroism in resonant metasurfaces with monoclinic lattices}
\end{center}

\section{Numerical simulations}

\begin{figure}[h]
	\centering
	\includegraphics[width=0.55\linewidth]{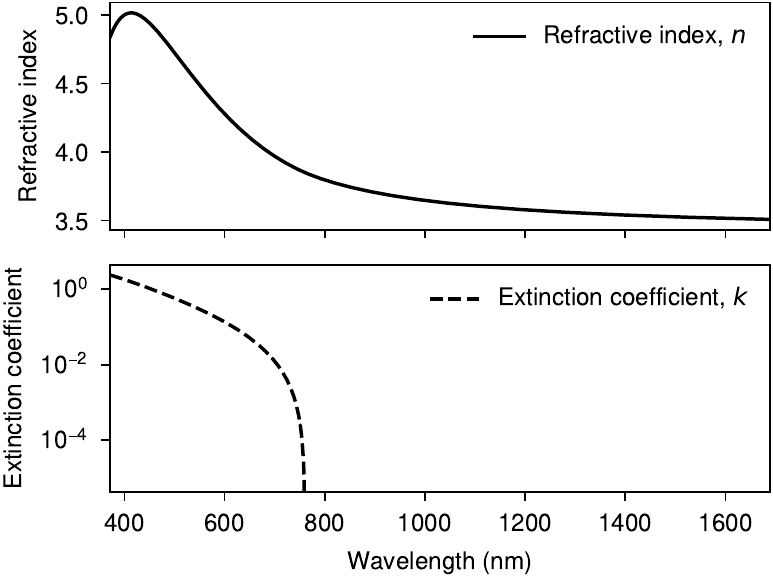}
	\caption{Refractive index and extinction coefficient of Si extracted from ellipsometry data.}
	\label{fig:nkSi}
\end{figure}

All numerical simulations were performed in the Wave Optics module of COMSOL Multiphysics.
The near-field distributions, resonant wavelengths, and $Q$-factors are simulated using the eigenfrequency solver.
Linear and nonlinear transmission simulations are simulated in the frequency domain.
Metasurface was placed on a semi-infinite substrate surrounded by a perfectly matched layer mimicking an infinite region in the vertical direction.
The simulation area is the unit cell with Floquet periodic boundary conditions which simulate an infinite size of the metasurface in a transverse plane.
The dispersion of the refractive index of Si is extracted from the ellipsometry data (Fig.~\ref{fig:nkSi}), while that of SiO$_2$ is taken from Refs.~\cite{Malitson1965JOSA,Polyanskiy2024SciData}.

The background field is set manually via custom code using Fresnel equations.
The third harmonic generation (THG) is calculated in the undepleted pump approximation using the domain polarization feature~\cite{COMSOL_SHG_example}. The nonlinear polarization current is calculated as $P_{i}^{(3\omega)} = \varepsilon_0 \hat{\chi}^{(3)}_{ijkm} E_{j}^{(\omega)} E_{k}^{(\omega)} E_{m}^{(\omega)}$, 
where $\hat{\chi}^{(3)}$ tensor has 21 nonzero elements based on the Si symmetry class $m3m$ (227-th space group)~\cite[Table~1.5.4]{boyd2008nonlinear}:
\begin{align}
	xxxx &= yyyy = zzzz, \nonumber \\
	yyzz &= zzyy = zzxx = xxzz = xxyy = yyxx, \nonumber  \\
	yzyz &= zyzy = zxzx = xzxz = xyxy = yxyx, \nonumber  \\
	yzzy &= zyyz = zxxz = xzzx = xyyx = yxxy.
\end{align}
Here $ijkm$ is the shorthand for $\hat{\chi}^{(3)}_{ijkm}$ for $i,j,k,m = x,y,z$.
Among 21 nonzero elements only 4 are independent but for simplicity we set these to be equal.
We assume that the crystallographic axes are aligned with the
metasurface grating direction and incident field direction, i.e. with the base Cartesian unit vectors $(\mathbf{\hat{x}}, \mathbf{\hat{y}}, \mathbf{\hat{z}})$.
Once the total fields are calculated for the RCP and LCP background fields, $\mathbf{E}_{^{\text{R}} _{\text{L}}}$, the  complex transmission amplitude coefficients are calculated as $t^{(n\omega)}_{^{\text{RR}} _{\text{LL}} } = \braket{\mathbf{\hat{e}}_{\pm}}{\mathbf{E}_{^{\text{R}} _{\text{L}}}^{(n\omega)}} =  \frac{1}{A} \iint\limits_{A} \mathbf{\hat{e}}^{*}_{\pm} \cdot \mathbf{E}_{^{\text{R}} _{\text{L}}}^{(n\omega)} (x,y,z_0) \dd x \dd y$, where $A$ is the area of the $z=z_0$ plane located at the edge of the simulation area from the opposite side of excitation, and $\mathbf{\hat{e}}_{\pm} = (\mathbf{\hat{x}} \pm \iu \mathbf{\hat{y}} )/\sqrt{2}$ are the unit vectors in the circuar polarization basis.
Integration over surface $A$ averages the output signal over the angles, so it gives only the $0$-th diffraction order.
Finally, the transmission coefficients are calculated as $T^{(\omega)}_{^\text{RR} _{\text{LL}}} = \frac{n_{\text{subs}}}{n_{\text{host}}} \abs{t_{^\text{RR} _{\text{LL}}}^{(\omega)}}^2$, and the output harmonic intensity is $I_{^\text{RR} _{\text{LL}}}^{(3\omega)} \propto \abs{t_{^\text{RR} _{\text{LL}}}^{(3\omega)}}^2$, where the proportionality coefficient is unimportant within the scope of this work.

\section{Numerical analyses of the nonlinear chiral characteristics}

\begin{figure}[h!]
	\centering
	\includegraphics[width=1\linewidth]{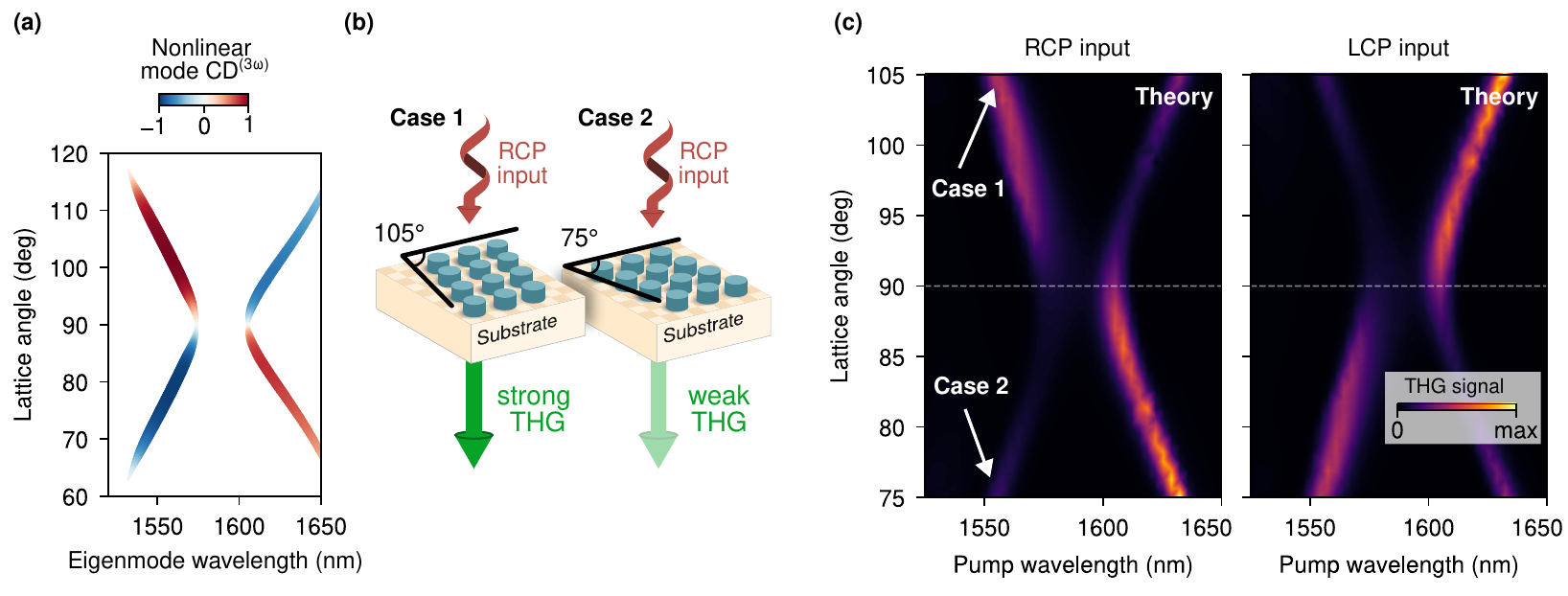}
	\caption{\textbf{Nonlinear theory}.
		\textbf{(a)} Mode nonlinear circular dichroism calculated via coupled mode theory for the excitation from the top.
		\textbf{(b)} Example of two metasurfaces with opposite handedness (enantiomers). Once the mode of the excitation wavelength has a high chirality parameter it is coupled better with RCP input rather than LCP input or vice versa. Once the coupling strength ($\abs{m_{n\text{R}}}$ or $\abs{m_{n\text{L}}}$) is high for a particular input, this leads to a higher energy transfer into the third harmonic generation (THG) signal.
		The chessboard pattern is a guide for the eye to show a rectangular lattice. 
		\textbf{(c)} THG signal as a function of the lattice angle and the pump wavelength. For a rectangular lattice, i.e. $\varphi = 90^{\circ}$, THG signals are identical for both input polarization. However, the change of the lattice angle below or above $90^{\circ}$ one can see that one branch is brighter for one inclination and darker for another. This directly correlates with the eigenmode analyses made in Fig.~\eqref{fig:concept} from the main text.
	}
	\label{fig:nonlinear_theory}
\end{figure}

In Fig.~\ref{fig:nonlinear_theory} we vary the lattice angle in the same manner as in Fig.~\ref{fig:concept}(b) from the main text and examine the THG signal in transmission collecting only the $0$-th diffraction order. 
Fig.~\ref{fig:nonlinear_theory}{(a)} shows the  nonlinear mode $\mathrm{CD}^{(3\omega)}$ for the excitation from the top, while
Fig.~\ref{fig:nonlinear_theory}{(b)} demonstrates that for a chiral mode, the asymmetry in coupling strength  with different light polarizations ($\abs{m_{n\text{R}}}$ or $\abs{m_{n\text{L}}}$) manifests itself in the nonlinear signal as well. 
This leads to a stronger interaction with the preferred polarization, resulting in a higher intensity of the third-harmonic generation signal. The figure visually depicts this effect for different lattice angles.
One can see a clear correspondence of the nonlinear mode CD and third harmonic signal intensities (compare panels (a) and {(c)} in Fig.~\ref{fig:nonlinear_theory}). 

\section{Common reasons of the mismatch between theoretical and experimental results}
\label{sec:mismatch_reasons}

There are several common reasons of quantitative differences between the numerical simulations data and experimental data, which include
\begin{enumerate}
	\item \textbf{Deviation of excitation shape from a plane wave}. In practice, the excitation beam has a shape of a Gaussian beam. The finite size of the Gaussian beam waist corresponds to the contribution of multiple incident waves with oblique $\mathbf{k}$-vectors, in contrast to common theoretical models accounting only for excitation by a single plane wave propagating normally. As a result, the measured spectrum represents a spectral convolution in the range of angles of incidence, instead of a response at a singular  $\mathbf{k}$-vector. Due to spectral averaging, the observed linewidth appears smaller. This effect can be reduced by using an excitation objective with a low numerical aperture.
	\item \textbf{Contribution of structural imperfections to the value of $Q$-factor}. Standard numerical models provide the value of the $Q$-factor of an ideal infinite structure determined solely by radiation and material absorption losses, $\Im(\varepsilon) \neq 0$. The resonant $Q$-factor of realistic structures is affected by additional factors, originating from various energy loss channels~\cite[Sec. 2.3]{koshelev2022PhDthesis}\cite{Hermann1984,Koshelev2020Science}. In case of relatively small losses, the total $Q_{\text{real}}$ can be decomposed into partial contributions as
	\begin{equation}
		Q^{-1}_{\text{real}} \approx Q^{-1}_{\text{ideal}}  + Q^{-1}_{\text{imp}}, 
	\end{equation}
	where $Q_{\text{ideal}}$ is the $Q$-factor of the ideal infinite structure, while $Q_{\text{imp}}$ comes from all possible practical imperfections, including surface roughness-induced scattering off imperfections, structural disorder and losses from the sample edges due to its finite size.
	
	Quantitative calculation of $Q_{\text{imp}}$ represents a numerical challenge. More recent experimental studies on fabrication robustness show that for resonant Si metasurfaces operating in the visible and near-infrared spectral ranges, the typical value of $Q_{\text{imp}}$ is in the range $100\text{--}200$~\cite{kuhne2021fabrication}. 
	
	We note that for the increase of the nonlinear signal value, it is favorable to achieve the so-called critical (optimal) coupling regime, when $Q_{\text{ideal}} \approx Q_{\text{imp}}$~\cite{Hermann1984, Maier2006OE,seok2011radiation}. In this case, the local electric field inside the metasurface material reaches its maximal value, which increases generated nonlinear currents, and, consequently, the emitted harmonic signal~\cite{koshelev2019nonlinear}, that we can collect in the far-field.
	
	\item \textbf{Limited detection range of spectrometer}. The effect of limited spectrometer detection range on the value of circular dichroism can be verified experimentally. For this, we can verify the identity of equality of circular dichroism from two different input directions, $\mathrm{CD}_{\text{co}}(\omega) = \mathrm{CD}^{\prime}_{\text{co}}(\omega)$, that follows from the electromagnetic reciprocity~\cite[Ch.~9]{Shalin2024AllDielectricNanophotonics}. Figure~2c of the main text shows the comparison of $\mathrm{CD}_{\text{co}}$ and $\mathrm{CD}^{\prime}_{\text{co}}$. One can see deviations from  $\mathrm{CD}_{\text{co}}(\omega) = \mathrm{CD}^{\prime}_{\text{co}}(\omega)$ in the spectral range of input wavelengths higher than $1600~\text{nm}$, which signifies a lower measurement accuracy beyond that point.
	
\end{enumerate}

\section{Anti-crossing of the modes 1 and 2}

\begin{figure}[h]
	\centering
	\includegraphics[width=0.4\linewidth]{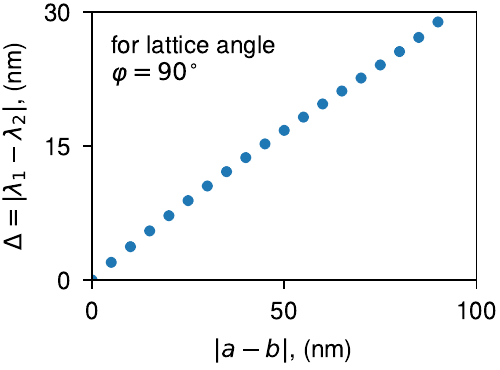}
	\caption{Results of the numerical calculations for the $\Delta = |\lambda_1 - \lambda_2|$ as the function of the $|a - b|$, where $\lambda_{1,2}$ are the eigenmode  wavelengths, and $a$, $b$ are the oblique lattice constants. }
	\label{fig:Delta_vs_a_minus_b}
\end{figure}

Transition from the cubic ($a = b$, $\varphi = 90^{\circ}$) to rectangular lattice ($a \neq b$, $\varphi = 90^{\circ}$) lowers the symmetry of the system. This removes the degeneracy of the modes 1 and 2 as they start to transform under the same irreducible representation. This and similar transformations can be studied in terms of the perturbation theory~\cite{Igoshin2024PRB,Muljarov2018OL,CanosValero2024PhysRevRes},which lies out of the scope of this work. Here we limit ourselves to the empirical observation that the anti-crossing gap is linearly proportional to $|a - b|$, so that
\begin{equation}
	\Delta \propto |a - b|,
\end{equation}
where $\Delta = |\lambda_1 - \lambda_2|$ is the difference of the eigenmode wavelengths, see Fig.~\ref{fig:Delta_vs_a_minus_b}.

\section{Background field}

We can analyze the problem within the framework of the scattered field, expressed via the Lippmann-Schwinger equation as a convolution of the metasurface’s Green’s function and the background field. The total electric field is given by
\begin{equation}
	\vb{E}^{(\omega)} = \vb{E}^{(\omega)}_{\text{bg}} + \vb{E}^{(\omega)}_{\text{sc}}
	\label{eq
	}
\end{equation}
where $\vb{E}^{(\omega)}_{\text{bg}}$ is the Fresnel field, which includes incident, reflected, and transmitted fields. Considering the perturbation of the permittivity $\varepsilon(\vb{r}) = \varepsilon_{\text{bg}}(\vb{r}) + \Delta \varepsilon(\vb{r})$, we employ {the dyadic Green's function decomposition into resonant states} to calculate the total field. This method involves using the Mittag-Leffler expansion of the Green's function into pole contributions, allowing us to find modes of a complex system through the modes of an analytically solvable system via a set of infinite matrix equations~\cite{Weiss2018PRB}.

\section{Chiral transmission characteristics}

S-matrix problem in terms of the vectors of amplitudes of incoming $\bf a$ and outgoing $\bf b$ circularly polarized waves is formulated as (Fig.~\ref{fig:channels}a): 
\begin{equation}\label{Sxyvec}
	{\bf b}={\bf S}\cdot{\bf a},
\end{equation}
or, expressed by the components:
\begin{equation}\label{SLReq}
	\begin{bmatrix}
		b_R \\
		b_L \\
		b'_R \\
		b'_L
	\end{bmatrix}=
	\begin{bmatrix}
		r_{RR} & r_{RL} & t'_{RR} & t'_{RL}\\
		r_{LR} & r_{LL} & t'_{LR} & t'_{LL}\\
		t_{RR} & t_{RL} & r'_{RR} & r'_{RL}\\
		t_{LR} & t_{LL} & r'_{LR} & r'_{LL}
	\end{bmatrix}
	\begin{bmatrix}
		a_R \\
		a_L \\
		a^{\prime}_R \\
		a^{\prime}_L
	\end{bmatrix}
\end{equation}
where the indexes of the reflection and transmission coefficients denote the final and initial circular polarizations respectively, and the prime describes the back (substrate) metasurface side. For metasurfaces made of reciprocal materials, the S-matrix in such notation is symmetric.

Optical rotation ($\mathrm{OR}$) and circular dichroism ($\mathrm{CD}$), conventionally used to characterize optical chirality of light transmission, are defined as:
\begin{equation}
	\label{OR}
	\mathrm{OR} = \frac{1}{2}(\arg t_{LL} - \arg t_{RR}),
\end{equation}
and
\begin{equation}
	\label{CDco}
	\mathrm{CO}_{\text{co}} = \frac{|t_{RR}|^2 - |t_{LL}|^2}{|t_{RR}|^2 + |t_{LL}|^2}.
\end{equation}
One can also introduce a large number of characteristics of other chiral light transformations described by Eq.~\eqref{SLReq}, such as, for example, the conversion circular dichroism: 
\begin{equation}
	\label{CDcross}
	\mathrm{CD}_{\text{cross}} = \frac{|t_{RL}|^2 - |t_{LR}|^2}{|t_{RL}|^2 + |t_{LR}|^2}.
\end{equation}
Note that due to reciprocity $\mathrm{CO}_{\text{co}} $  is independent of the metasurface side while $\mathrm{CD}_{\text{cross}}$ inverts its sign when the sides are switched.

For third harmonic generation (THG), one introduces the third harmonic circular dichroism (TH-CD) (similarly to Ref.~\cite{Petralli-Mallow1993JPhysChem}): 
\begin{equation}
	\label{THCD}
	\mathrm{CD}^{(3\omega)} = \frac{I_{R}^{(3\omega)} - I_{L}^{(3\omega)}}{I_{R}^{(3\omega)} + I_{L}^{(3\omega)}},
\end{equation}
where $I_{R,L}^{(3\omega}$ are the TH intensities, when RCP or LCP wave is used for pumping. Generally, it \textit{does depend} on the metasurface side. 

\section{Chiral CMT for linear and nonlinear transmission}\label{sec:CMT}

\begin{figure}
	\centering
	\includegraphics[width=0.8\linewidth]{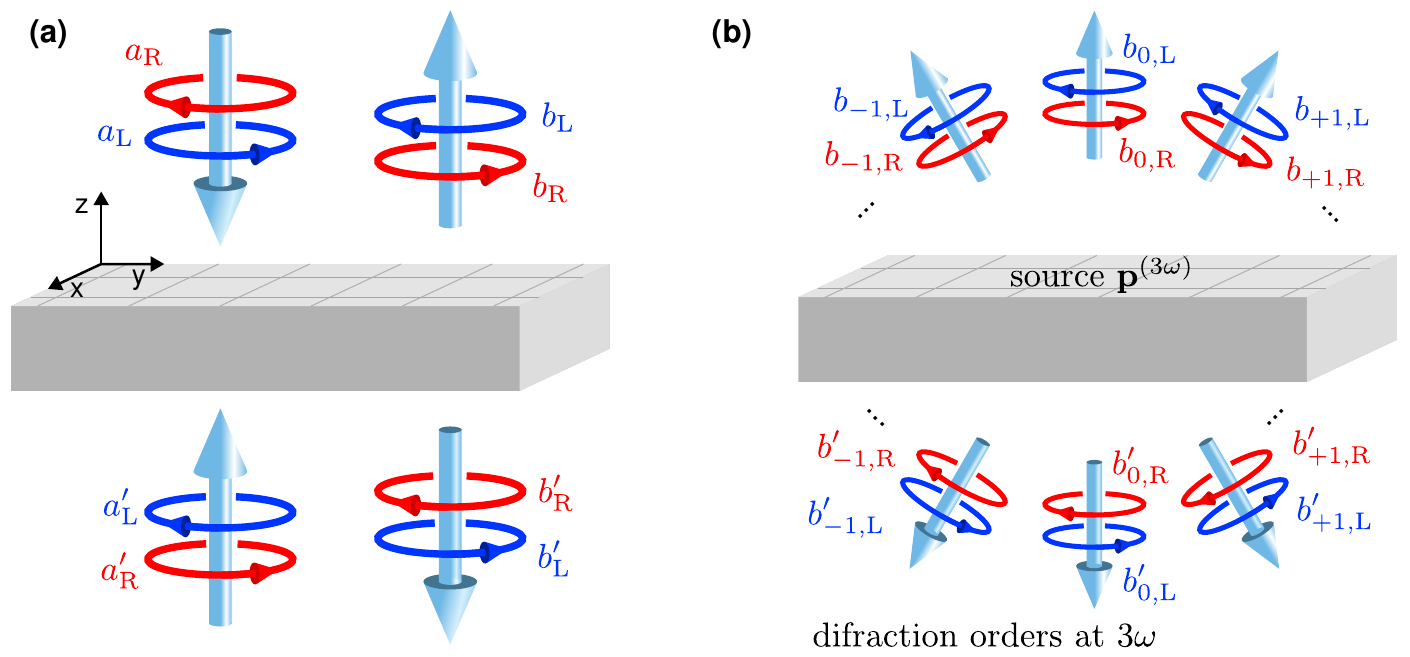}
	\caption{Illustration of the \textbf{(a)} input, $\vb{a}$, and output, $\vb{b}$, channels in a circular polarization basis at the input frequency $\omega$ and \textbf{(b)} output channels at the third harmonic frequency, $3\omega$. Since the metasurface has a 2D in-plane periodicity, strictly speaking, we have to introduce two indices of diffraction orders. But in favor of clearness in \textbf{(b)} we show only one set of diffraction orders.}
	\label{fig:channels}
\end{figure}

In the coupled-mode theory (CMT) the incident electromagnetic radiation oscillating with a real frequency $\omega$ excites 2 relevant for us eigenstates, each of them having a slow-varying complex amplitude $p_1$ and $p_2$ (the fast time dependence is assumed to be $\eu^{-\iu \omega t}$ and it is omitted for the sake of compactness). 
This can be easily generalized for an arbitrary number of modes~\cite[Ch.~9]{Shalin2024AllDielectricNanophotonics}.
The vector of  amplitudes $\mathbf{p} = [p_1,p_2]^T$ describes the state time evolution as an oscillator driven by the incident waves:
\begin{equation}\label{CMT1}
	\frac{\dd \vb{p}}{\dd t} =  (\vb{\iu\Omega} - \vb{\Gamma}) \cdot \vb{p} + \vb{M}^T \cdot \vb{a}. 
\end{equation} 
where the matrix
\begin{equation}\label{M}
	{\bf M}=
	\begin{bmatrix}
		m_{1R} & m_{2R}\\
		m_{1L} & m_{2L}\\
		m^{\prime}_{1R} & m^{\prime}_{2R}\\
		m^{\prime}_{1L}& m^{\prime}_{2L}\\ 
	\end{bmatrix}.
\end{equation}
describes the coupling of the eigenstates to RCP or LCP waves incident on the corresponding metasurface side 
~\cite{Weiss2018PRB,Gorkunov2020Aug,koshelev2022PhDthesis,Koshelev2020Science}:
\begin{equation}
	m_{n\text{R,L}} = A_n \int \limits_{\text{meta}} \Delta \varepsilon (\vb{r}) \vb{E}_n (\omega_n, \vb{r}) \cdot \vb{E}_{\text{bg}}^{(\mathrm{R}, \mathrm{L})} (\omega_n, \vb{r}) \dd^{3} \vb{r}
	\label{eq:m}
\end{equation}
with the eigenmode electric field $\vb{E}_n$ and the background electric field of RCP or LCP wave $\vb{E}_{\text{bg}}^{(\mathrm{R},\mathrm{L})}$ including the Fresnel fields reflected and transmitted by the substrate without the metasurface~\cite[Sec.~VI]{Weiss2018PRB}. 
$\Delta \varepsilon (\vb{r})$ denotes the difference between the permittivities of the metasurface and the background which in our case includes the substrate.
The primed coupling parameters describe the coupling to the waves incident on the back metasurface side. 
The $A_n$ is the normalization coefficient. A symmetric magnetic counterpart of Eq.~\eqref{eq:m} is absent due to the non-magnetic nature of the materials, i.e., $\Delta \mu (\vb{r}) =0$.
It follows from the scattering formalism for resonant states~\cite{Weiss2018PRB} that one should substitute the complex eigenmode frequency $\omega_n$ into the background field.

For non-interacting eigenstates one can write the diagonal matrices:
\begin{equation}\label{OmegaGamma}
	{\bf \Omega}=	
	\begin{bmatrix}
		\omega-\omega_1  &  0\\
		0 & \omega-\omega_2  
	\end{bmatrix},\ 
	{\bf \Gamma}=	
	\begin{bmatrix}
		\gamma_1  & 0\\
		0 & \gamma_2  
	\end{bmatrix}.
\end{equation}
with $\omega_{1,2}$ being the real eigenstate frequencies and $\gamma_{1,2}$ being the decay rates. 

The irradiation of outgoing waves is described by another CMT equation:
\begin{equation}\label{CMT2}
	{\bf b}={\bf M}{\bf p}+{\bf C}{\bf a}, 
\end{equation}
where the coupling matrix ${\bf M}$ is the same as in \eqref{CMT1} to ensure reciprocity.
The last term in Eq.~\eqref{CMT2} describes a non-resonant background transmission. If it is achiral and polarization-independent, we can describe it by the matrix:  
\begin{equation}\label{Cmatr}
	{\bf C}=
	\begin{bmatrix}
		0 & \rho & \tau & 0\\
		\rho & 0 & 0 & \tau\\
		\tau & 0 & 0 & \rho\\
		0 & \tau & \rho & 0
	\end{bmatrix}.
\end{equation}

A stationary solution of Eqs.~\eqref{CMT1} reads as
\begin{equation} \label{eq:p_sol}
	\mathbf{p} = - (\iu\mathbf{\Omega} -  \mathbf{\Gamma})^{-1}\mathbf{M}^{T} \mathbf{a},
\end{equation}
and, according to \eqref{CMT2}, it determines the S-matrix as:
\begin{equation}\label{SmatCMT}
	{\bf S}={\bf C}-{\bf M}\cdot (\iu\mathbf{\Omega} -  \mathbf{\Gamma})^{-1} \cdot{\bf M}^T.
\end{equation}
In the absence of dissipation, the matrix should remain  unitary, ${\bf S}^\dag {\bf S}=1$, which, in particular, relates the coupling parameters with the radiation decay rates as~\cite{Fan2003JOSAA}:
\begin{equation}\label{gammarad}
	2\gamma_n^{\rm rad}=|m_{nR}|^2+|m_{nL}|^2+|m^{\prime}_{nR}|^2+|m^{\prime}_{nL}|^2.
\end{equation}

Following  Eq.~\eqref{SmatCMT}, the corresponding transmission coefficients are explicitly expressed as:
\begin{equation}\label{tRR}
	t_{RR}=t'_{RR}=\tau-\sum_{n=1,2} \frac{m^{\prime}_{nR}m_{nR}}{\iu(\omega-\omega_n)-\gamma_n},
\end{equation} 
\begin{equation}\label{tLL}
	t_{LL}=t'_{LL}=\tau-\sum_{n=1,2} \frac{m^{\prime}_{nL}m_{nL}}{\iu (\omega-\omega_n)-\gamma_n},
\end{equation} 
\begin{equation}\label{tRL}
	t_{RL}=t'_{LR}=-\sum_{n=1,2} \frac{m^{\prime}_{nR}m_{nL}}{\iu(\omega-\omega_n)-\gamma_n},
\end{equation} 
\begin{equation}\label{tLR}
	t_{LR}=t'_{RL}=-\sum_{n=1,2} \frac{m^{\prime}_{nL}m_{nR}}{\iu(\omega-\omega_n)-\gamma_n}.
\end{equation}

For THG, the waves irradiation at the tripled frequency $3\omega$ can be described by an analog of Eq.~\eqref{CMT2} without a background term (as there are no such incident waves):
\begin{equation}\label{CMTTH}
	{\bf b}^{(3\omega)}={\bf M}^{(3\omega)}{\bf p}^{(3\omega)}, 
\end{equation}
where ${\bf p}^{(3\omega)}$ is the vector of amplitudes of states with eigenfrequencies around $3\omega$ and ${\bf M}^{(3\omega)}$ is the matrix of their coupling to free-space waves. Note that the set of amplitudes ${\bf b}^{(3\omega)}$ usually includes several orders of diffraction (Fig.~\ref{fig:channels}b), as metasurfaces are rarely subwavelength for the waves at $3\omega$, so it has the form of
\begin{equation} \label{eq:M3w}
	\mathbf{M}^{(3\omega)} = \begin{pmatrix}
		\vdots  & \vdots  & & \vdots \\ 
		\mathbf{m}_{1,-1} & \mathbf{m}_{2,-1} & & \mathbf{m}_{N,-1} \\
		\mathbf{m}_{1, 0} & \mathbf{m}_{2, 0} & \dots & \mathbf{m}_{N,0}  \\
		\mathbf{m}_{1,+1} & \mathbf{m}_{2,+1} & & \mathbf{m}_{N,+1} \\ 
		\vdots & \vdots & & \vdots
	\end{pmatrix}, \quad \text{with} \quad \mathbf{m}_{jm} = \begin{pmatrix}
		m^{(3\omega)}_{jm,\text{R}}  \vspace{0.2cm} \\  
		m^{(3\omega)}_{jm,\text{L}} \vspace{0.2cm} \\
		m^{\prime(3\omega)}_{jm,\text{R}} \vspace{0.2cm} \\
		m^{\prime(3\omega)}_{jm,\text{L}} 
	\end{pmatrix},
\end{equation}
and 
\begin{equation}
	\mathbf{p}^{(3\omega)} = \begin{pmatrix}
		p_1^{(3\omega)} \\ p_2^{(3\omega)} \\ \vdots \\ p_N^{(3\omega)} 
	\end{pmatrix}.
\end{equation}

Each state ${p}_j^{(3\omega)}$ from the set ${\bf p}^{(3\omega)}$ is excited due to intrinsic coupling with the states $p_n$ from the set ${\bf p}$ as:
\begin{equation}\label{eq:p_3w}
	\frac{\dd {p}_j^{(3\omega)}}{\dd t } = [\iu(3\omega-\omega_j)- \gamma_j] p_j^{(3\omega)} + X_{jik\ell} p_{i}p_{k}p_{\ell}, 
\end{equation}
where Einstein's summation rule is assumed in the later term, and where $X_{jik\ell}$ are the nonlinear coupling parameters arising due to the nonlinear susceptibility of metasurface material $\hat{\chi}^{(3)}$~\cite[Appendix A]{Koshelev2024JOPT}:
\begin{equation}
	X_{jik\ell} 
	\propto \int \limits_{\text{meta}}  \mathbf{E}^{(3\omega)}_{j} (\mathbf{r}) \cdot  \hat{\chi}^{(3)} \left( \Shortstack{. . .} \right)   \mathbf{E}_{i} (\mathbf{r}) \mathbf{E}_{k} (\mathbf{r}) \mathbf{E}_{\ell } (\mathbf{r}) \dd^3 \mathbf{r}.
	\label{eq:X}
\end{equation}
We notice that writing Eq.~\eqref{eq:p_3w}, we assume that the THG is dominated by resonant channels, which is a good approximation for the considered case of high-Q modes at the pumping frequency.

In the stationary THG regime, 
\begin{equation}
	p_{j}^{(3\omega)} = - \iu \frac{X_{jik\ell} p_{i}p_{k}p_{\ell}}{\omega_j - 3\omega - \iu \gamma_j},
\end{equation}
and for a particular $\nu$-th output channel from the set ${\bf b}^{(3\omega)}$ one can write 
\begin{equation}
	b_{\nu}^{(3\omega)} = M_{\nu j}^{(3\omega)} p_j^{(3\omega)} = \tilde{M}^{(3\omega)}_{\nu i k \ell} p_i p_k p_{\ell},
\end{equation}
where $p_i$ are given by Eq.~\eqref{eq:p_sol} and coefficients
\begin{equation}
	\tilde{M}^{(3\omega)}_{\nu i k \ell} = - \iu  \sum \limits_{j}\frac{M^{(3\omega)}_{\nu j} X_{jik\ell} }{ \omega_j - 3\omega - \iu \gamma_j}
\end{equation}
are introduced. Each coefficient $\tilde{M}^{(3\omega)}_{\nu i k \ell}$ is independent of the input polarization at the pump frequency which is emphasized by the $(3\omega)$ superscript. It contains many competing terms as long as the frequency $3\omega$ does not approach a solitary high-$Q$ resonance. Note that such resonances are very unlikely to appear in the diffraction spectral range.

We note that it is possible to introduce a \textit{full nonlinear scattering matrix}, which takes into account the reverse process~\cite{Achouri2018IEEE}. However, since the reverse process of $(3\omega)\to (\omega)$ is highly in-efficient, we omit such terms in the full nonlinear scattering matrix and consider only $(\omega) \to (3\omega)$ process.

\section{Linear and nonlinear circular dichroism}

If the quality factors of eigenstates around the pumping frequency are high enough, they contribute to the transmission separately. Then, substituting the coefficients (\ref{tRR}--\ref{tLR}) into \eqref{CDco}, one can write for frequencies close to $\omega_n$:
\begin{equation}
	\label{CDn}
	\mathrm{CD}_{\text{co}} (\omega \approx \omega_n) \approx \frac{|\tau- \frac{m^{\prime}_{nR}m_{nR}}{i(\omega-\omega_n)-\gamma_n}|^2 - |\tau- \frac{m^{\prime}_{nL}m_{nL}}{i(\omega-\omega_n)-\gamma_n}|^2}{|\tau- \frac{m^{\prime}_{nR}m_{nR}}{i(\omega-\omega_n)-\gamma_n}|^2 + |\tau- \frac{m^{\prime}_{nL}m_{nL}}{i(\omega-\omega_n)-\gamma_n}|^2}.
\end{equation}
Generally, its frequency dependence  is rather complex.  
In the particular case of absent background co-polarized transmission, $\tau=0$, this CD becomes frequency-independent and coincides with the eigenstate CD:
\begin{equation}
	\label{def_CD_n}
	\mathrm{CD}_{\text{co}} (\omega\approx\omega_n, \tau \approx 0) = 
	\mathrm{CD}_n = \frac{|m^{\prime}_{nR}m_{nR}|^2 - |m^{\prime}_{nL}m_{nL}|^2}{|m^{\prime}_{nR}m_{nR}|^2 - |m^{\prime}_{nL}m_{nL}|^2}.
\end{equation}
The conversion CD, in turn, always lacks pronounced spectral anomalies in the resonant ranges and  equals 
\begin{equation}
	\label{res_CD_cross}
	\mathrm{CD}_{\text{cross}} (\omega \approx \omega_n)= \frac{|m^{\prime}_{nR}m_{nL}|^2 - |m^{\prime}_{nL}m_{nR}|^2}{|m^{\prime}_{nR}m_{nL}|^2 - |m^{\prime}_{nL}m_{nR}|^2}.
\end{equation}

When the pumping frequency is close to the eigenfrequency $\omega_n$ of one high-$Q$ state, the output intensity of a $\nu$-th TH channel  depends on the side and handedness of the pumping:
\begin{equation}
	I^{(3\omega)}_{n,\text{R}} = \left(|\tilde{M}^{\prime (3\omega)}_{\text{R}}|^2 + |\tilde{M}^{\prime(3\omega)}_{\text{L}}|^2 \right) \frac{|a_{\text{R}}|^6|m_{n \text{R}}|^6}{[(\omega-\omega_n)^2+{\gamma}_n^2]^3},	\qquad 
	I^{(3\omega)}_{n,\text{L}}= \left(|\tilde{M}^{\prime(3\omega)}_{\text{R}}|^2 + |\tilde{M}^{\prime(3\omega)}_{\text{L}}|^2 \right)\frac{|a_{\text{L}}|^6|m_{n\text{L}}|^6}{[(\omega-\omega_n)^2+{\gamma}_n^2]^3},	
	\label{eq:I3w_top}
\end{equation}
for one side, and  
\begin{equation}
	I^{\prime (3\omega)}_{n,\text{R}}= \left(|\tilde{M}^{ (3\omega)}_{\text{R}}|^2 + |\tilde{M}^{ (3\omega)}_{\text{L}}|^2 \right)\frac{|a_{\text{R}}|^6|m^{\prime}_{n \text{R}}|^6}{[(\omega-\omega_n)^2+{\gamma}_n^2]^3},	\qquad 
	I^{\prime (3\omega)}_{n,\text{L}} = \left(|\tilde{M}^{ (3\omega)}_{\text{R}}|^2 + |\tilde{M}^{(3\omega)}_{\text{L}}|^2 \right) \frac{|a_{\text{L}}|^6|m^{\prime}_{n\text{L}}|^6}{[(\omega-\omega_n)^2+{\gamma}_n^2]^3},	
	\label{eq:I3w_bot}
\end{equation}
for the other. Here coefficient $\tilde{M}_{\text{R}}$ is equal to  $\tilde{M}_{\text{R}} = - \iu \sum_j m^{(3\omega)}_{j,0, \text{R}} X_{j,1,1,1} \frac{1}{\omega_j - 3\omega - \iu \gamma_j}$ (see Eq.~\eqref{eq:M3w}), once we consider only $0$-th diffraction order, see Fig.~\ref{fig:channels}b. 

In the approximation of one distinct resonance at the pump frequency, $\omega \approx \omega_n$, based on the definition Eq.~\eqref{THCD}, the TH-CD for different sides equals to
\begin{equation}\label{THCD1}
	\mathrm{CD}^{(3\omega)} (\omega \approx \omega_n) = \frac{|m_{nR}|^6-|m_{nL}|^6}{|m_{nR}|^6+|m_{nL}|^6}, \qquad 
	\mathrm{CD}^{\prime (3\omega)} (\omega \approx \omega_n) = \frac{|m^{\prime}_{nR}|^6-|m^{\prime}_{nL}|^6}{|m^{\prime}_{nR}|^6+|m^{\prime}_{nL}|^6}.
\end{equation}
We emphasize that, in general, $\mathrm{CD}^{(3\omega)} \neq \mathrm{CD}^{\prime (3\omega)}$.

\begin{figure}
	\centering
	\includegraphics[width=0.6\linewidth]{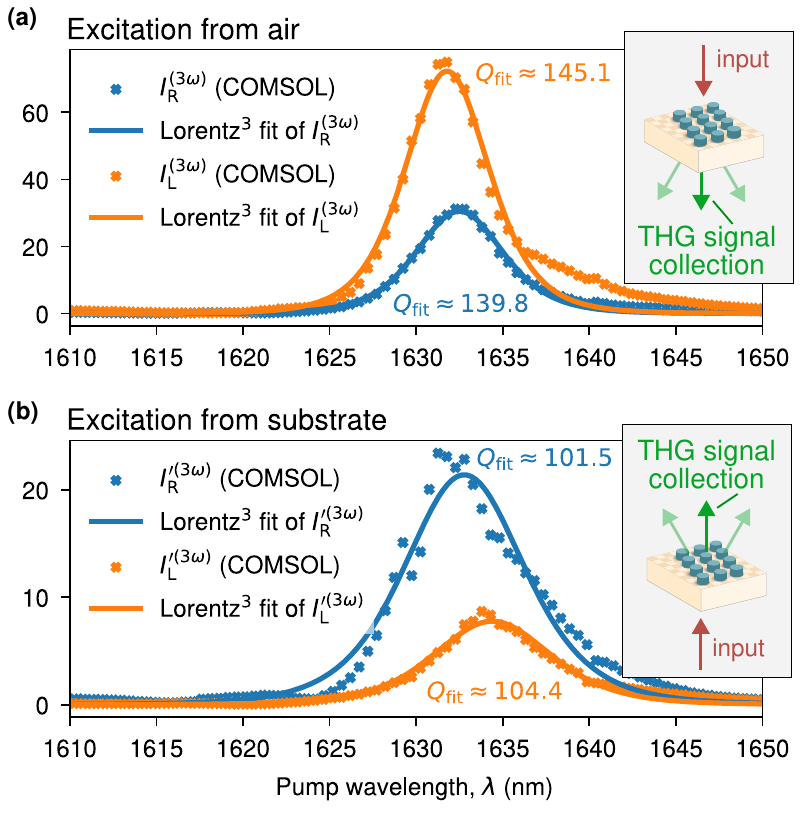}
	\caption{Resonance fit of the simulated THG signal for the excitation from air (a) and from the substrate (b). COMSOL eigenvalue solver predicts the $Q$-factor to be around $Q_2\approx 160$, and the fit based on the Eqs.~(\ref{eq:I3w_top}--\ref{eq:I3w_bot}) produces similar values.}
	\label{fig:THG_theory_top_down}
\end{figure}

To explicitly test the CMT applicability, we fit the TH intensity spectra obtained using COMSOL with analytical Eqs.~(\ref{eq:I3w_top}--\ref{eq:I3w_bot}) and observe excellent quantitative agreement, see Fig.~\ref{fig:THG_theory_top_down}. Moreover, the fitting allows us to resolve the eigenstate quality factor $Q_2$ which can be independently obtained from a solution of the eigenstate problem in COMSOL. We obtain although different but similar values of $Q_2$ between 100 to 140 for the four THG spectra, while the eigenstate solver predicts $Q_2\approx160$. This indicates remarkable accuracy of the taken approximations, as a complex nonlinear-optical process of THG indeed can be reproduced by relatively simple analytics.

\section{Experimental methods}

The design parameters were $a = 1100$~nm, $b=1000$~nm, $R = 430$~nm, $H = 400$~nm, and $\varphi = 75$~deg, see Fig.~\ref{fig:linear}(a) from the main text. 
The samples were fabricated using a combination of electron beam lithography, lift-off processes, and reactive ion etching (RIE). An amorphous silicon film with a thickness of 400~nm was deposited on a SiO$_2$ substrate by electron beam evaporation (Syskey Tech) at a rate of 0.5 \r{A} s$^{-1}$. An 80~nm PMMA film was then spin-coated onto the Si film and baked at 180$^{\circ}$C for one hour. Nanostructures were realized in PMMA by patterning it in an electron beam aligner (Raith E-line, 30 kV) and developing in MIBK/IPA solution for 30~s at $0 ^{\circ}$C. Afterwards, a 26~nm Cr film was directionally deposited by electron beam evaporation (0.3 \r{A} s$^{-1}$, Syskey Tech), and the nanostructures were transferred onto Cr by removing the PMMA residue in Remover PG for 8 h. The nanostructures were further transferred onto the silicon film by an etching process with inductively coupled plasma (Plasmalab System 100 ICP180). Then the Si metasurface was finally obtained after removing the Cr mask in chromium etchant for 10 min.

To compare with our theoretical predictions shown in Fig.~\ref{fig:linear}(d) from the main text we measure the transmission of RCP and LCP near-IR light through the metasurface. 
The incoming and outgoing circular light polarizations are set by polarizers and quarter wave plates in front of and behind the sample. 
The recorded spectra including two co-polarized (LL and RR) and two cross-polarized (LR and RL) transmittances are shown in Fig.~\ref{fig:linear}(c) from the main text. 
The co-polarized spectra demonstrate peaks that shifted related to each other. Similar behaviour is observed for cross-polarized spectra. 
We expected the second resonant mode near 1700~nm, however, due to the low sensitivity of the spectrometer {beyond 1600~nm} the resonant mode is non-pronounced in the recorded spectra. 
The calculated from the experimental data CD according to Eq.~\eqref{eq:CDco} from the main text and is shown in Fig.~\ref{fig:linear}(c). 
Overall, linear CD exhibits a value close to $0$, and changes from $-0.2$ to $+0.2$ near the resonance wavelength. 
The measured spectra is not perfectly matched with the theory due to the reasons discussed in Sec.~\ref{sec:mismatch_reasons}.

To investigate the resonant properties of THG generated from the metasurface, we pump it by a femtosecond laser with wavelengths in the range of 1500--1730~nm with a 1~nm step from two sides: from the air and from the substrate. The recorded THG spectra for RCP and LCP excitation show two distinct peaks corresponding to the two chiral modes studied above, see them compared with the theoretical THG spectra in Figs.\ref{fig:THG_theory_top_down}(b,c). In all cases, THG signals demonstrate resonant enhancement near two wavelengths corresponding to the resonant modes observed in the linear regime. 
Next, we evaluate the experimental nonlinear CD according to Eq.~\eqref{eq:NCDtot} from the main text as seen on the top panels of Figs.~\ref{fig:THG_theory_top_down}(b,c) from the main text. Being driven by the resonances, nonlinear CD varies from $-0.64$ to $0.32$ for different pump wavelengths.

\end{document}